\documentclass[aps,pre,amsmath,amssymb,twocolumn,showpacs]{revtex4}
\usepackage{graphics}
\usepackage{graphicx}
\usepackage{color}
\usepackage{relsize}

\def\be{\begin{equation}}
\def\ee{\end{equation}}

\def\ba{{\bf a}}

\def\bg{{\bf g}}
\def\bk{{\bf k}}
\def\br{{\bf r}}
\def\bu{{\bf u}}
\def\bv{{\bf v}}
\def\bx{{\bf x}}
\def\by{{\bf y}}

\def\bD{{\bf D}}

\def\lb{\label}

\def\bar{\overline}

\def\bdot{\hbox{\boldmath $\cdot$}}

\def\bzed{\hbox{\boldmath $0$}}
\def\grad{\hbox{\boldmath $\nabla$}}

\begin{document}


\title{Do Sweeping Effects Suppress Particle Dispersion in Synthetic Turbulence?}


\author{Gregory L. Eyink${\,\!}^{1,2}$}
\email{eyink@jhu.edu}
\author{Damien Benveniste${\,\!}^2$}
\affiliation{%
${\,\!}^{1}$Department of Applied Mathematics \& Statistics\\
and ${\,\!}^{2}$Department of Physics \& Astronomy\\
The Johns Hopkins University, USA}%


\date{\today}

\begin{abstract}
Synthetic models of Eulerian turbulence like so called ``Kinematic Simulations'' (KS) are often used as computational shortcuts for studying Lagrangian properties of turbulence. These models have been criticized by Thomson \& Devenish (2005), who argued on physical grounds that sweeping decorrelation effects suppress pair dispersion in such models.  We derive analytical results for Eulerian turbulence modeled by Gaussian random fields, 
in particular for the case with zero mean velocity. Our starting point is an exact integrodifferential equation for the particle pair separation distribution obtained from the Gaussian integration-by-parts identity. When memory times of particle locations are short, a Markovian approximation leads to a Richardson-type diffusion model.  We obtain a time-dependent pair diffusivity tensor of the form $K_{ij}({\bf r},t)=S_{ij}({\bf r})\tau(r,t)$ where $S_{ij}({\bf r})$ is the structure-function tensor and $\tau(r,t)$  is an effective correlation time of velocity increments.  Crucially, this is found to be the minimum value of three times: the intrinsic turnover time $\tau_{eddy}(r)$ at separation $r$, the overall evolution time $t,$ and the  sweeping time $r/v_0$  with $v_0$ the rms velocity.  
We study the diffusion model numerically by a Monte Carlo method.  With moderate inertial-ranges like those achieved in current KS, our model is found to reproduce the $t^{9/2}$ power-law for pair dispersion predicted by Thomson \& Devenish and observed in the KS. However, for much longer ranges, our model exhibits three distinct pair-dispersion laws in the inertial-range: a Batchelor $t^2$-regime, followed by a Kraichnan-model-like $t^1$ diffusive regime, and then a $t^6$ regime.  Finally, outside the inertial-range, there is another $t^1$ regime with particles  undergoing independent  Taylor diffusion. These scalings  are exactly the same as those predicted by Thomson \& Devenish for KS with large mean velocities, which we  argue hold also for KS with zero mean velocity.  Our results support the basic conclusion of Thomson \& Devenish (2005) that sweeping effects make Lagrangian properties of KS fundamentally different from hydrodynamic turbulence for very extended inertial-ranges.
\end{abstract}


\maketitle

\section{Introduction}\lb{intro}

How particle pairs separate in a turbulent flow has been a central subject of turbulence research
since the classical work of Richardson \cite{Richardson26}. Unfortunately, the phenomenon has proved
quite difficult to investigate by numerical solution of the fluid equations and by controlled laboratory 
experiments, especially because of the very large Reynolds numbers required. Many studies have therefore  
employed ``synthetic turbulence''  or ensembles of random velocity fields with some of the scaling 
properties of real turbulent velocities but which can be efficiently sampled even for very long scaling ranges.
For example, papers \cite{ElliottMajda96, FungVassilicos98, MalikVassilicos99, DavilaVassilicos03,
NicolleauYu04} have followed this approach and have reported substantial agreement of their numerical
simulations with the predictions of Richardson, including the famous ``$t^3$-law'' for the growth in time 
of mean square pair separation distances. 

The validity of these results has been called into question, however. A paper of Chaves et al. 
\cite{Chavesetal03} pointed out that the  use of synthetic turbulence to model Eulerian velocity statistics 
implies sweeping effects of large-scale eddies on particle motions that diverge with the Reynolds number.
Those authors suggested to employ synthetic ensembles such as Gaussian random fields 
to model instead the turbulent statistics of Lagrangian velocities. In a simple one-dimensional 
Gaussian model of Eulerian velocities they found analytically that large-scale sweeping effects 
``localized'' particle pairs and prevented them from separating. Subsequently, in a very interesting 
paper \cite{ThomsonDevenish05}, Thomson \& Devenish have proposed an intuitive picture 
how sweeping affects particle dispersion in synthetic models of Eulerian turbulence. The key point
is that large-scale eddies in real turbulence advect both particles and smaller scale eddies, while 
large-scale eddies in synthetic turbulence advect only particle pairs and not smaller eddies. 
This fact implies that particle pairs at separations $r$ in synthetic turbulence should experience 
rapidly changing relative velocities, as they are swept into new, statistically independent eddies.
This occurs on a ``sweeping'' time-scale $\tau_{sw}(r)\sim r/v_0,$ where $v_0$ is the rms velocity set 
by the largest eddies in the synthetic ensemble. Thomson \& Devenish assume a diffusion process of 
pair separations with an eddy-diffusivity $K(r)\sim \delta u^2(r)\tau_{sw}(r)$ and $\delta u^2(r)$ the 
mean-square relative velocity at separation $r.$ In an ensemble with Kolmogorov scaling 
$\delta u^2(r)\sim (\varepsilon r)^{2/3},$ this yields $dr^2/dt\sim K(r) \sim \varepsilon^{2/3}r^{5/3}/v_0$ 
and the solution
\be \langle r^2(t)\rangle \sim \frac{\varepsilon^4 t^6}{v_0^6}. \lb{t6} \ee   
Note that this implies considerably slower growth than Richardson's $t^3$-law \footnote{The two laws 
can be written as $<\!\!r^2(t)\!\!>_{{\rm Rich}} \sim L^2 (t/t_L)^3$ and $<\!\!r^2(t)\!\!>_{{\rm TD}} 
\sim L^2 (t/t_L)^6$ in terms of the velocity integral scale $L$ and the large-eddy turnover time $t_L\sim
L/v_0,$ by using $\varepsilon\sim v_0^3/L.$
In their regime of validity $t<t_L,$ one has $<\!\!r^2(t)\!\!>_{{\rm TD}}\,\, \ll \,\,<\!\!r^2(t)\!\!>_{{\rm Rich}}.$}.
Thomson \& Devenish argued for the above prediction in the case of a large mean sweeping, with 
$v_0$ replaced by the mean speed $\bar{u}.$ In the case of a zero-mean velocity ensemble, they 
argued instead for a $t^{9/2}$-growth law, intermediate between $t^3$ and $t^6$ (see our section 
\ref{predict} below). These predictions were supported in \cite{ThomsonDevenish05} by the numerical 
technique of ``Kinematic Simulations'' (KS) \cite{ElliottMajda96, FungVassilicos98, MalikVassilicos99, 
DavilaVassilicos03,NicolleauYu04}. The previous contrary numerical results were explained on 
various grounds, e.g. the use of an adaptive time-stepping scheme in \cite{ElliottMajda96} which 
did not resolve the small sweeping time $\tau_{sw}(r)$ and its effect on particle dispersion.  
 
The issues raised by the paper of Thomson \& Devenish have still not been fully resolved. The 
numerical simulations in \cite{ThomsonDevenish05} used another form of adaptive time-stepping, which 
was suggested in \cite{Osborneetal06} to be responsible for the observation of a $t^{9/2}$ growth.
Thomson \& Devenish then repeated their simulations with a fixed small time-step and reported 
the same $t^{9/2}$ law \cite{DevenishThomson09}. The most recent simulations of Nicolleau \& 
Nowakowski \cite{NicolleauNowakowski11} for their longest scaling ranges show some evidence 
of the Thomson-Devenish sweeping effects, but the reported scaling laws are intermediate between 
those of Richardson and of Thomson-Devenish and agree with neither theory. Thus, there is still 
considerable uncertainty in the literature regarding the validity of the Thomson-Devenish theory. The 
question is important, because synthetic turbulence is a useful testing ground for numerical and theoretical 
methods, and because comparison of particle dispersion in synthetic and real turbulence 
illuminates the physical mechanisms of the latter. 

Because of the disagreement of the numerical simulations of different groups, it is useful to have 
analytic results. The Thomson-Devenish arguments apply to a wide array of synthetic turbulence 
models, but Gaussian velocity ensembles are the most mathematically tractable. We therefore 
consider here the use of Gaussian random fields as models of turbulent Eulerian velocities. 
More precisely, we take the advecting velocity field $\bu(\bx,t)$ to be a Gaussian random field 
with mean $\bar{\bu}(\bx,t)$ and covariance $C_{ij}(\bx,t;\by,s)=  \langle u_i'(\bx,t) u_j'(\by,s)\rangle$ 
for the fluctuations $\bu'=\bu-\bar{\bu}.$ Specific models of interest are similar to those studied 
in \cite{Chavesetal03}, with $\bar{\bu}(\bx,t)=\bar{\bu}$ independent of space and time and 
with covariance defined for $0<\alpha<2, 0<\beta<2$ by 
\begin{eqnarray}
&& C_{ij}(\bx,t;\by,s) \cr
&& \,\,\,\,\,\,\,\,\,\,
=D_2 \int d^dk\,\, e^{-D_3 k_L^\beta |t-s|} \frac{e^{i\bk\bdot(\bx-\by)}}{k_L^{d+\alpha}} P_{ij}(\bk). 
\lb{velcov} 
\end{eqnarray} 
Here $k_L^2=k^2+1/L^2$ and $P_{ij}(\bk)=\delta_{ij}-k_ik_j/k^2$ is the projection onto the 
subspace of ${\mathbb R}^d$ orthogonal to $\bk.$ The Gaussian random field $\bu(\bx,t)$ so 
defined is statistically homogeneous in space, stationary in time, and solenoidal. The length
$L$ is proportional to the integral length-scale. The scaling properties of the model at scales
smaller than $L$ are similar to those of real turbulence. For example, the single-time covariance 
for $r\ll L$ is calculated to be
\begin{eqnarray} 
&& C_{ij}(\bx,t;\by,t) \sim D_0L^\alpha \cr
&&   \,\,\,\,\,\,\,\,\,\, - D_1 r^\alpha [(d+\alpha-1)\delta_{ij}-\alpha \hat{r}_i\hat{r}_j]
     + O(r^2/L^2) 
\lb{powlawcov} \end{eqnarray} 
with $\br=\bx-\by.$ See \cite{EyinkXin00}, p.686. Kolmogorov 1941 dimensional scaling 
corresponds to the exponents $\alpha=\beta=2/3.$ We shall consider also in this paper 
Gaussian velocity models whose energy spectra coincide with KS models which 
have been studied numerically \cite{ThomsonDevenish05,Osborneetal06,
DevenishThomson09,NicolleauNowakowski11}.  The incompressibility of these models 
will be used in an essential way, although much of our analysis
applies to more general models, e.g. with any degree of compressibility. 

The principal results of this paper are as follows. For a general Gaussian model 
of Eulerian turbulence we carefully derive the diffusion approximation for pair 
dispersion assumed in the argument of Thomson-Devenish \cite{ThomsonDevenish05},
under the assumption of short memory times for particle locations. We furthermore 
obtain a closed formula, eq.(\ref{Gauss-KLform}), for the 2-particle eddy-diffusivity in a general 
Gaussian model. For the specific models with covariance (\ref{velcov}) we obtained more 
explicit results, which, under the conditions $\alpha<1$ and either $\beta<1$ or frozen turbulence
with $D_3=0,$ verify the Thomson-Devenish argument about sweeping decorrelation effects.
In particular, we obtain under these conditions a 2-particle eddy-diffusivity  tensor of the form 
$K_{ij}(\br,t)=S_{ij}(\br)\tau(r,t),$ where $S_{ij}({\bf r})$ is the structure-function tensor and 
$\tau(r,t)$  is an effective correlation time of velocity increments. Crucially, $\tau(r,t)$ is the minimum 
of the intrinsic turnover time $\tau_{eddy}(r)$ at separation $r$, the overall evolution time $t,$ 
and the sweeping time $r/v_0.$ Although this result confirms the sweeping decorrelation effect, 
we argue that the pair-dispersion law for zero mean-velocity ensembles at high Reynolds numbers 
is different from the $t^{9/2}$ suggested by Thomson \& Devenish \cite{ThomsonDevenish05}. 
Instead, we argue that there are distinct ranges of power-laws $t^2,$ $t^1,$ $t^6$ and then 
$t^1$ again at successively longer times, exactly as Thomson \& Devenish argued for 
ensembles with large mean velocities. We carry out careful numerical Monte Carlo 
simulations with our diffusion model which verify these behaviors in the model at very
high Reynolds numbers. We also present Monte Carlo results for our diffusion model 
at the moderate Reynolds numbers employed in current KS work, and reproduce then 
both the ``$t^3$-law'' and ``$t^{9/2}$-law'' that have been reported in KS at comparable 
Reynolds numbers. We thus argue that the current KS results in the literature are not yet 
probing asymptotic regimes and the true scaling in KS at very high Reynolds numbers 
will be the same as in our diffusion model.  
 
The detailed analytical derivation of diffusion models is presented in section \ref{Derive} of the 
paper, and predictions for their dispersion laws discussed in section \ref{predict}. Our numerical 
methods are described and validated in section \ref{num} , and then used to obtain results
for mean-square particle separations and other statistics. A concluding section \ref{Conclude}
briefly discusses the results.

\section{Derivation of the Diffusion Model}\lb{Derive}

In this section we present the derivations of our main analytical results.  A reader 
who is only interested in physical conclusions and not the detailed justifications 
may skip to our final formula (\ref{Gauss-KLform}) for the pair-diffusivity and the 
following discussion. 

\subsection{Gaussian Integration-by-Parts Identity}\lb{intpart}

We show first that the transition probability of particle pairs in Gaussian velocity 
ensembles obeys an exact evolution equation, as a consequence of the well-known
integration-by-parts identity or Donsker-Furutsu-Novikov relation (see \cite{Frisch95}, 
section 4.1). Let $\bu(\bx,t)$ be the random turbulent velocity field and let 
the fluid particle position that satisfies
\be {{d}\over{dt}}\bx(t)=\bu(\bx,t), \,\,\,\, \bx(t_0)=\ba \lb{A1} \ee
be denoted as $\bx_\bu(\ba,t_0|t)$, or $\bx(\ba,t)$ for short.
Define the ``fine-grained PDF'' of 2-particle positions as 
\be P_{2,\bu}(\bx_2,\bx_1,t|\ba_2,\ba_1,t_0) = \prod_{n=1}^2
\delta^d(\bx_n-\bx_\bu(\ba_n,t_0|t)). \lb{A2} \ee
Then the PDF of 2-particle positions is given by 
\be P_2(\bx_2,\bx_1,t|\ba_2,\ba_1,t_0) = 
\langle P_{2,\bu}(\bx_2,\bx_1,t|\ba_2,\ba_1,t_0)\rangle, \lb{A5} \ee
where the average is over the random velocity field $\bu.$ 

Taking the time-derivative of (\ref{A2}) and using (\ref{A1}) it is a calculus 
exercise to show that 
\be \partial_t  P_{2,\bu}(t)  =-\sum_{n=1}^2 \grad_{\bx_n}\bdot\left[ 
(\bar{\bu}(\bx_n,t)+\bu'(\bx_n,t))P_{2,\bu}(t)\right], \lb{B1} \ee
where the velocity has been decomposed into its mean and fluctuating part
$\bu(\bx,t)=\bar{\bu}(\bx,t)+\bu'(\bx,t).$
The average of the second term on the righthand side can be obtained using 
Gaussian integration-by-parts \cite{Frisch95} 
\begin{eqnarray}
 \langle u'_i(\bx,t))P_{2,\bu}(t)\rangle& = & \int d^dy\int ds \,\,C_{ik}(\bx,t;\by,s) \cr
&& \,\,\,\,\,\,\times
\left\langle \frac{\delta}{\delta u_k(\by,s)}P_{2,\bu}(t)\right\rangle,
\end{eqnarray}  
where $C_{ij}(\bx,t;\by,s)=\langle u_i'(\bx,t) u_j'(\by,s)\rangle.$ To represent 
the functional derivative we introduce the Lagrangian response function
\be G_{ij}(\ba,t;\by,s)\equiv \frac{\delta x_i(\ba,t)}{\delta u_j(\by,s)}, 
\lb{Lagrespfn} \ee
so that 
\be
 \frac{\delta}{\delta u_k(\by,s)}P_{2,\bu}(t) = \sum_{m=1}^2 
-\partial_{x^j_m}P_{2,\bu}(t)\cdot G_{jk}(\ba_m,t;\by,s). \ee
The result of averaging (\ref{B1}) is the drift-diffusion equation 
 \begin{eqnarray}
& & \partial_t  P_{2}(t)  = -\sum_{n=1}^2 \grad_{\bx_n}\bdot
\left[\bar{\bu}^*(\bx_n,t)P_{2}(t)\right] \cr
& & +\sum_{n,m=1}^2\partial_{x^i_n}\partial_{x^j_m}\left[D_{ij}(\bx_n,\bx_m,t,t_0)
P_{2}(t)\right]. 
\lb{A12} \end{eqnarray}    
with
\be \bar{\bu}^*(\bx,t)=\bar{\bu}(\bx,t)+\left.\partial_{x^j}D_{ij}(\bx,\bx',t,t_0)\right|_{\bx'=\bx}
\lb{A13} \ee
the mean velocity plus a fluctuation-induced drift, and  with the diffusivity tensor
\begin{eqnarray}
&& D_{ij}(\bx_n,\bx_m,t,t_0) \equiv  \int_{t_0}^t ds \int d^dy \,\,C_{ik}(\bx_n,t;\by,s) \cr
&& \,\,\,\,\,\,\,\,\,\,\,\,\,\,\,\,\,\,\,\,\,\,\,\,
\times \langle G_{jk}(\ba_m,t;\by,s)|\bx_2,\bx_1,t;\ba_2,\ba_1,t_0\rangle
\lb{D-gauss} \end{eqnarray}                                             
where
\begin{eqnarray}
&& \langle G_{jk}(\ba_m,t;\by,s)|\bx_2,\bx_1,t;\ba_2,\ba_1,t_0\rangle \cr 
&& \,\,\,\,\,\,\,\,\,\,\,\,\,\,\,\,\,\,\,\,\,\,\,\,\,\,\,\,\,\,\,\,\,\,\,\,\,\,\,\,
=\frac{\langle G_{jk}(\ba_m,t;\by,s) P_{2,\bu}(t)\rangle}{P_2(t)} 
\end{eqnarray}
is the conditional average of the response function given that the two particles
start in locations $\ba_2,\ba_1$ at time $t_0$ and end up  at locations 
$\bx_2,\bx_1$ at time $t.$

We now develop a more useful expression for the response function (\ref{Lagrespfn}).  
It is straightforward to show by functional differentiation of the equation of motion
(\ref{A1}) that 
\be \partial_t G_{ij}=\frac{\partial u_i}{\partial x_k}(\bx(\ba,t_0|t),t)G_{kj}
              +\delta_{ij} \delta^d(\by-\bx(\ba,t_0|t))\delta(t-s). \ee 
This equation may be solved as
\be G_{ij}(\ba,t;\by,s)=\left\{\begin{array}{ll}
                                                g_{ij}(\by,s|t)\delta^d(\by-\bx(\ba,t_0|s)) & t>s>t_0 \cr
                                                0 & {\rm o.w.} \cr
                                                \end{array} \right. \lb{Gg} \ee
with $\bg(\by,s|t)={\rm Texp}\left(\int_s^t dr \, \frac{\partial\bu}{\partial\bx}(\bx(\ba,t_0|r),r)\right)$
the time-ordered exponential matrix for the trajectory which satisfies $\bx(\ba,t_0|s)=\by.$ This
notation is made natural by an alternative derivation of (\ref{Gg}) based on the flow composition 
identity
\be \bx(\ba,t_0|t)= \bx(\bx(\ba,t_0|s),s|t). \lb{comp} \ee 
Taking the functional derivative $\delta/\delta u_j(\by,s)$ of (\ref{comp}) and using the chain rule gives
\be \frac{\delta x_i(\ba,t)}{\delta u_j(\by,s)}=
\left.\frac{\partial x_i}{\partial y_k}(\by,s|t)\right|_{\by=\bx(\ba,s)}
\frac{\delta x_k(\ba,s)}{\delta u_j(\by,s)}. \ee
On the other hand, it is readily seen that 
the functional derivative of the integral form of the particle equation 
of motion (\ref{A1}), 
gives
\be \frac{\delta x_k(\ba,s)}{\delta u_j(\by,s)} =\delta_{jk} \delta^d(\by-\bx(\ba,t_0|s))
\theta(s-t_0). \ee
Thus, eq.(\ref{Gg}) is rederived with $g_{ij}(\by,s|t)=\frac{\partial x_i}{\partial y_j}(\by,s|t).$
If (\ref{Gg}) is substituted into the formula (\ref{D-gauss}) it yields
\begin{eqnarray}
&& D_{ij}(\bx_n,\bx_m,t,t_0) \equiv  \int_{t_0}^t ds \int d^dy_m \,\,C_{ik}(\bx_n,t;\by_m,s) \cr
&& \,\,\,\,\,\,\,\,\,\,\,\,\,\,\,\,\,\,\,\,\,\,\,\,
\times \langle g_{jk}(\by_m,s|t)|\bx_2,\bx_1,t;\by_m,s;\ba_2,\ba_1,t_0\rangle \cr
&& \,\,\,\,\,\,\,\,\,\,\,\,\,\,\,\,\,\,\,\,\,\,\,\,
\times P(\by_m,s|\bx_2,\bx_1,t;\ba_2,\ba_1,t_0)
\lb{D-gauss-2} \end{eqnarray}             
where $P(\by_m,s|\bx_2,\bx_1,t;\ba_2,\ba_1,t_0)$ is the conditional probability density 
of the position of particle $m$ at time $s$ given the positions of both particles at times $t$
and $t_0.$ This formula for the diffusivity when substituted into (\ref{A12}),(\ref{A13}) 
gives the final form of our exact evolution equation for the 2-particle transition probability.  

\subsection{Markovian Approximation}\lb{mark}

Despite appearances, the evolution in the exact equation (\ref{A12}) is non-Markovian in general.
It is clear from formula (\ref{D-gauss-2}) that the 2-particle diffusion matrix is a function not only of the 
particle positions $\bx_1,\bx_2$ at time $t,$ but also of the positions $\ba_1,\ba_2$ at time $t_0.$
This dependence was suppressed in our notations, but the evolution, in principle, retains a long-time 
memory of the initial conditions. Only in special cases can the evolution be shown to be Markovian. 
The famous example is the Gaussian velocity field that is delta-correlated in time, the so-called
{\it Kraichnan model} \cite{Kraichnan68,Falkovichetal01}, for which 
\be C_{ik}(\bx,t;\by,s) = C_{ik}(\bx,\by;t) \delta(t-s). \ee
Substituting into (\ref{D-gauss-2}) and using 
\be g_{jk}(\by_m,t|t)=\delta_{jk} \lb{gtt} \ee
and
\be P(\by_m,t|\bx_2,\bx_1,t;\ba_2,\ba_1,t_0)=\delta^d(\by_m-\bx_m) \lb{Ptt} \ee
gives (with the ``$\frac{1}{2}$ delta-function rule'' for the upper limit of integration)
$$D_{ij}(\bx_n,\bx_m,t,t_0)=\frac{1}{2} C_{ij}(\bx_n,\bx_m,t). $$
Thus, in this case rigorously there is no dependence of the diffusion matrix $\bD$ on 
$\ba_1,\ba_2$ and the well-known diffusion model is  obtained \cite{Kraichnan68,Falkovichetal01}.
Another example with Markovian particle evolution is the velocity field obtained as the superposition 
of Gaussian random wave trains with very high frequencies, so that the group velocity of the waves 
greatly exceeds the root-mean-square velocity \cite{Balk02}. This example has direct relevance to 
KS simulations with ``eddy-turnover frequency'' $\omega_n=\lambda\sqrt{k_n^3 E(k_n)}$ in the 
limit $\lambda\gg 1$ of large ``unsteadiness'' parameter. 

The description as a diffusion should generally hold reasonably well if the correlation time of the 
Gaussian velocity field is short enough, since the integrand in (\ref{D-gauss-2}) then becomes negligible 
at values of $s<t$ for which there is sizable dependence on $\ba_1,\ba_2$. With this motivation, we make the 
{\it Markovian approximation}
\begin{eqnarray}
&& D_{ij}(\bx_n,\bx_m,t,t_0) \equiv  \int_{t_0}^t ds \int d^dy_m \,\,C_{ik}(\bx_n,t;\by_m,s) \cr
&& \,\,
\times \langle g_{jk}(\by_m,s|t)|\bx_2,\bx_1,t;\by_m,s\rangle P(\by_m,s|\bx_2,\bx_1,t). 
\lb{D-gauss-M} \end{eqnarray}    
The physical assumption is that for times ordered as $t_0\ll s<t$ the position of the particle at time $s$
is determined mainly by its position at time $t$ and is negligibly dependent on the position at the initial 
time $t_0$. The worst case for this approximation is clearly the ``frozen velocity'' model with infinite correlation 
time, when times $s\gtrsim t_0$ in the integral are not suppressed by decay of correlations. Such $s$
values give an undamped contribution also in general for times $t-t_0$ much smaller than the velocity 
correlation time. However, it is easy to check that  the exact result (\ref{D-gauss}) 
[or (\ref{D-gauss-2})] and the Markovianized result (\ref{D-gauss-M}) both give 
\be \frac{d}{dt}D_{ij}(\bx_n,\bx_m,t,t_0) = C_{ij}(\bx_n,t;\bx_m,t) +O(t-t_0) \ee
so that, for $t-t_0$ much smaller than the correlation time, 
\be D_{ij}(\bx_n,\bx_m,t,t_0)= C_{ij}(\bx_n,t_0;\bx_m,t_0)(t-t_0) + O((t-t_0)^2). \lb{tshort-M} \ee 
Thus the Markovian approximation becomes exact in this limit. We note in passing that the 
Kraichnan-Lundgren theory of 2-particle dispersion \cite{Kraichnan66,Lundgren81} when applied 
to the Gaussian velocity ensemble gives a result almost identical to the formula (\ref{D-gauss-M})
(for more discussion, see \cite{Eyink12}).

The formula (\ref{D-gauss-M}) from the Markovian approximation can be further simplified. It is intuitively 
clear that conditioning on the location of both particles is superfluous in an average of a random variable
that involves only one of these particles. In fact, it can be easily established from the definitions 
(\ref{A2}),(\ref{A5}) that 
\begin{eqnarray} 
P(\by,s|\bx',\bx,t) &= & \int d^dy' \,P(\by',\by,s|\bx',\bx,t) \cr
                              &=& \int d^dy' \,\langle \delta^d(\by'-\bx(\bx',t|s))\delta^d(\by-\bx(\bx,t|s)) \rangle \cr
                              &=& \langle \delta^d(\by-\bx(\bx,t|s))\rangle = P(\by,s|\bx,t). 
\end{eqnarray}                              
A similar argument gives 
\be  \langle g_{jk}(\by,s|t)|\bx',\bx,t;\by,s\rangle =
\langle g_{jk}(\by,s|t)|\bx,t;\by,s\rangle. \ee       
More generally, we may define the PDF
\begin{eqnarray} 
&& P(\bg,t;\by',\by,s|\bx',\bx,t)= \cr
&& 
\langle \delta^{d\times d}(\bg-\bg(\by,s|t)) \delta^d(\by'-\bx(\bx',t|s))\delta^d(\by-\bx(\bx,t|s)) \rangle \cr
&&
\end{eqnarray}   
and mimic the previous argument to show that
\be    P(\bg,t;\by,s|\bx',\bx,t)=  P(\bg,t;\by,s|\bx,t).   \ee
Then  
\begin{eqnarray}
P(\bg,t|\by,s;\bx',\bx,t) &=&  \frac{P(\bg,t;\by,s|\bx',\bx,t)}{P(\by,s|\bx',\bx,t)}   \cr      
                                      &=&  \frac{P(\bg,t;\by,s|\bx,t)}{P(\by,s|\bx,t)}   \cr 
                                      &=& P(\bg,t|\by,s;\bx,t).
\end{eqnarray}
It follows from these facts that
\begin{eqnarray}
&& D_{ij}(\bx_n,\bx_m,t,t_0) \equiv  \int_{t_0}^t ds \int d^dy_m \,\,C_{ik}(\bx_n,t;\by_m,s) \cr
&& \,\,\,\,
\times \langle g_{jk}(\by_m,s|t)|\bx_m,t;\by_m,s\rangle P(\by_m,s|\bx_m,t),
\lb{D-gauss-4} \end{eqnarray}    
which is the final form of the Markovian approximation for the diffusion tensor. 

We now consider the special case when the velocity field is statistically homogeneous 
in space. In that case, the drift velocity in (\ref{A13}) is independent of $\bx$ and simplifies 
to $\bar{\bu}^*(t)=\bar{\bu}(t),$ due to homogeneity and incompressibility \footnote{This is
the first point where we have invoked incompressibility.}. Furthermore, a 
simplified equation can be derived for the transition probability of the 2-particle 
separation vector $\br=\bx_2-\bx_1,$ defined by 
\begin{eqnarray}
P_2(\br,t|\br_0,t_0) &=& \int d^da \,\,  P_2(\bx+\br,\bx,t|\ba+\br_0,\ba,t_0) \cr
                              &=& \int d^da \,\,  P_2(\bx,\bx-\br,t|\ba+\br_0,\ba,t_0), \cr
                              \, &&\,
\end{eqnarray}                              
which is also independent of $\bx.$ Since the diffusion tensor $D_{ij}(\bx_n,\bx_m,t)$
depends only on the difference $\bx_n-\bx_m$ in the homogeneous case, the equation
(\ref{A12}) with the substitutions $\br=\bx_2-\bx_1$ and
\be \grad_{\bx_2}\longrightarrow \grad_\br, \,\,\,\, \grad_{\bx_1}\longrightarrow -\grad_\br,\ee  
yields the diffusion equation 
\be \partial_tP_2(\br,t|\br_0,t_0) = \partial_{r^i}\partial_{r^j}\left[K_{ij}(\br,t,t_0)P_2(\br,t|\br_0,t_0) \right],
\lb{Keq} \ee
with the eddy-diffusivity tensor
\begin{eqnarray}
&& K_{ij}(\br,t,t_0) \cr
&& =2D_{ij}(\bzed,\bzed,t,t_0)-D_{ij}(\br,\bzed,t,t_0)-D_{ij}(\bzed,\br,t,t_0) \cr
&& =\int_{t_0}^t ds \int d^dy \, S_{ik}(\br;\by,t,s)\cr
&& \,\,\,\,\,\,\,\,\,\,\,\,\,\,\,\,\,\,\,\,\,\,\,\,\,\,\,\,\,\, \times
 \langle g_{jk}(\by,s|t)|\bzed,t;\by,s\rangle P(\by,s|\bzed,t) \cr
&&  
\lb{exact2}\end{eqnarray} 
and we define the 2nd-order structure function of velocity increments at two points
$\bzed,\by$ and two times $t,s:$
\be S_{ik}(\br;\by,t,s)=\langle[u_i'(\br,t)-u_i'(\bzed,t)][u_k'(\by+\br,s)-u_k'(\by,s)]\rangle. 
\lb{2t2pS} \ee

If furthermore the velocity field is assumed to be statistically stationary in time, then we can 
take $t-t_0\longrightarrow t$ and $t_0\longrightarrow 0$, to obtain 
\be \partial_tP_2(\br,t|\br_0,0) = \partial_{r^i}\partial_{r^j}\left[K_{ij}(\br,t)P_2(\br,t|\br_0,0) \right],
\lb{Keq2} \ee
with 
\begin{eqnarray}
&& K_{ij}(\br,t) =\int_{-t}^0 d\tau  \int d^dy \, S_{ik}(\br;\by,0,\tau) \cr
&& \,\,\,\,\,\,\,\,\,\,\,
 \times \langle g_{jk}(\by,\tau|0)|\bzed,0;\by,\tau\rangle P(\by,\tau|\bzed,0)
\lb{Markov} \end{eqnarray} 
by the change of variables $\tau=s-t.$

\subsection{Structure Function and One-Particle Distribution Function}\lb{struc1pdf} 

The integral over $\by$ in the above formula (\ref{Markov}) converges at large $y$
because of decay in the two-point structure function and in the 1-particle transition probability.
Physically, rapid decay is due to the facts that increments separated by great distances 
are uncorrelated and particles have low probability to be swept to large distances.  
Both of these effects can be easily quantified.  

To evaluate the two-point structure function, we use a standard identity that expresses it 
in terms of the single-point 2nd-order structure function (\cite{YaglomMonin75}, p.102):
\begin{eqnarray}
&& S_{ik}(\br;\by,0,\tau) = \frac{1}{2}\left[ S_{ik}(\by+\br,0,\tau)+S_{ik}(\by-\br,0,\tau)\right.\cr
&& \,\,\,\,\,\,\,\,\,\,\,\,\,\,\,\,\,\,\,\,\,\,\,\,\,\,\,\,\,\,\,\,\,\,\,\,\,\,\,\,\,\,\,\,\,\,\,\,\,\,\,\,\,\,\,\,\,\,\,\,\,\,\,\,\,\,\,\,\left.
-2S_{ik}(\by,0,\tau)\right]. 
\lb{YM102} \end{eqnarray}
We first consider the single-time case with $\tau=0.$ For the spatial power-law covariance 
(\ref{powlawcov}) with  $0<\alpha<2$, the single-point structure 
function becomes
\begin{eqnarray} 
&& S_{ij}(\br)=\left.2\left[C_{ij}(\bzed,\tau)-C_{ij}(\br,\tau)\right]\right|_{\tau=0} \cr
&& \,\,\,\, \sim 2 D_1 r^{\alpha}\left[(d+\alpha-1)\delta_{ij}
     -\alpha \hat{r}_i\hat{r}_j\right]+O(r^2/L^2)
\lb{S2} \end{eqnarray}     
for $r\ll L.$ The formula (\ref{YM102}) implies in general that $S_{ik}(\br;\by,0,\tau=0)\sim S_{ik}(\br)$
for $y\ll r,$ whereas in the particular case (\ref{S2}) it gives 
\be S_{ik}(\br;\by,0,\tau=0)=O(r^2/y^{2-\alpha}) \lb{2tS2} \ee
for $r\ll y\ll L.$  When $y\gg L,$ there is generally exponential or fast power-law decay, depending on the precise 
assumptions about the fall-off of the spectrum at low $k.$ The 2-time structure function $S_{ij}(\br;0,\tau)$ 
shows a similar behavior as the single-time structure function, except that there is a new length 
$L_\beta(\tau)=(D_3|\tau|)^{1/\beta}$ with eddies smaller than this scale decorrelated by time $|\tau|.$ 
As seen from (\ref{velcov}), the decorrelation is associated to an exponential decay of the cospectrum, 
with $L_\beta(t)$ acting as an effective ``dissipation scale.'' 
Thus, $S_{ij}(\br;0,\tau)$ scales $\propto r^2$ for $r\ll L_\beta(\tau),$ while formula (\ref{S2}) holds for 
$L_\beta(\tau)\ll r\ll L.$ Thus, the decay law (\ref{2tS2}) is found when $\tau\neq 0$ only for the range of values
$\max\{r,L_\beta(\tau)\}\ll y\ll L$ and is limited to times $|\tau|<L^\beta/D_3.$ For $y\ll \max\{r,L_\beta(\tau)\}$
instead $S_{ik}(\br;\by,0,\tau)$ is independent of $\by$ and for $y\gg L$ the decay is again like that for $\tau=0.$

The 1-particle transition probability should be dominated by large-scale sweeping and thus 
have the form
\be P(\by,\tau|\bzed,0)=\frac{1}{(2\pi)^{d/2}v_0^d|\tau|^d}
      \exp(-|\by-\bar{\bu}\tau|^2/2v_0^2\tau^2) \lb{gauss-v} 
\ee
to a good approximation, with $v_0$ the root mean square velocity.  We hereafter consider 
mainly the case $\bar{\bu}=\bzed.$ For the Gaussian random field with mean zero and covariance 
(\ref{velcov}), $v_0\propto D_2L^\alpha.$ In that case, it has been verified by a formal scaling 
analysis in \cite{Chavesetal03}, section 7, that the leading-order motion of particles for large $L$ 
is indeed ballistic with a constant, random velocity $\bv=\by/\tau$ chosen from a Gaussian 
ensemble with rms value $v_0.$ Subleading corrections were also obtained in 
\cite{Chavesetal03} to account for the effects of the change of the velocity in space 
and time. Note that (\ref{gauss-v}) decays rapidly for $y\gg v_0|\tau|.$ 

\subsection{Stability Matrix}\lb{stab}

The most difficult term to evaluate in (\ref{Markov}) is the conditional average of the stability matrix
$\bg(\by,\tau|0).$ Existence of this matrix requires a short-distance cutoff $\eta$ on the ``inertial-range'' 
scaling behavior in the model covariance (\ref{velcov}) and (\ref{powlawcov}), which otherwise corresponds 
to velocity fields only H\"older continuous and non-differentiable in space. Even with the cutoff,  the matrix 
$\bg(\by,\tau|0)$ will grow exponentially in $|\tau|$ almost surely, with rate determined  by the leading 
Lyapunov exponent $\lambda \propto (D_1/\eta^{2-\alpha})^{1/2}.$ It is thus far from clear {\it a priori} that 
the conditional average even remains finite in the limit $\eta\rightarrow 0.$ 

We begin by evaluating this term for the ``frozen'' velocity field with infinite correlation time 
(or $D_3=0$ in eq.(\ref{velcov})).  A key observation here is that the Gaussian transition probability 
(\ref{gauss-v}) implies that particles are swept from point $\by$ to $\bzed$ in time $|\tau|$ along straight 
lines with a constant speed $v=y/\tau$ generally of order $v_0.$ The velocity-gradient field 
$\grad\bu(\bx,t)$ has a spatial correlation of order $\eta,$ so that the particle trajectories 
contributing in (\ref{Markov}) will see a constant in space but rapidly changing velocity-gradient 
with a correlation time $\sim \eta/v_0.$ Thus, one can expect that the Lagrangian velocity-gradient will 
be well approximated by the model of a Gaussian field that is delta-correlated  in time, for which the statistics 
of the stability matrix  has been much studied. 

To make this argument more formally, consider the spatial covariance of the 
velocity-gradient in the frozen case $C_{ij,mn}(\br)= \langle u_{i,m}'(\br) u_{j,n}'(\bzed)
\rangle,$ where $u_{i,m}'=\partial u_i'/\partial x_m.$ By twice differentiating (\ref{powlawcov}) 
and then averaging over the direction  of the unit vector $\hat{\br},$ it is calculated to be 
\begin{eqnarray}
&&  \bar{C}_{ij,mn}(r) = D'_1 r^{\alpha-2}\left[(d+1)\delta_{ij}\delta_{mn}\right. \cr
&& \,\,\,\,\,\,\,\,\,\,\,\,\,\,\,\,\,\,\,\,\,\,\,\,\,\,\,\,\,\,\,\,\,\,\,\,\,\,\,\,\,\,\,\,\,\,\,\,\,\,\,\,\,\,\,\,\,\,\,\,\,\,\,\,\,\,
\left.-(\delta_{im}\delta_{jn}+\delta_{in}\delta_{jm})\right] 
\lb{velgradcov}  \end{eqnarray} 
with $D'_1=\frac{D_1\alpha(\alpha-2)}{d}\left[\frac{\alpha-4}{d+2}+2+\frac{d}{\alpha-2}\right]>0$
for $d\geq 2$ and $0<\alpha<2.$ This covariance holds for $r>\eta,$ whereas the covariance 
for $r<\eta$ is essentially constant 
and can be taken to be given by (\ref{velgradcov}) with $r=\eta.$
A particle swept with velocity $v$ will see a random velocity-gradient with temporal
correlation obtained by substituting $r=vt$ in (\ref{velgradcov}).
Thus, the (Eulerian) velocity-gradients in a Lagrangian frame can be taken as Gaussian with 
covariance
\begin{eqnarray}
&&  \langle u_{i,m}'(t)u_{j,n}'(0)\rangle = D_1''\frac{\eta^{\alpha-1}}{v}  \delta_\eta(t) 
\left[(d+1)\delta_{ij}\delta_{mn}\right. \cr
&& \,\,\,\,\,\,\,\,\,\,\,\,\,\,\,\,\,\,\,\,\,\,\,\,\,\,\,\,\,\,\,\,\,\,\,\,\,\,\,\,\,\,\,\,\,\,\,\,\,\,\,\,\,\,\,\,\,\,\,\,\,\,\,\,\,\,
\left.-(\delta_{im}\delta_{jn}+\delta_{in}\delta_{jm})\right] 
\lb{velgradcov2}  \end{eqnarray} 
with $D_1''=2\left(\frac{2-\alpha}{1-\alpha}\right)D_1'$ and $\delta_\eta(t) = \frac{1}{t_\eta}\Delta(\frac{1}{t_\eta}),$ 
for $t_\eta=\eta/v$ and
\be \Delta(t)=  \frac{1-\alpha}{2(2-\alpha)} \times \left\{\begin{array}{ll}
                                  1 &  \mbox{for $|t|<1$} \cr
                                  t^{\alpha-2} & \mbox{for $|t|>1$} \cr
                                 \end{array} \right. . \ee
Since $\Delta(t)$ is integrable for $\alpha<1$ with 
$ \int_{-\infty}^{+\infty} dt\,\,\Delta(t) = 1,$ one then has $\lim_{\eta\rightarrow 0}\delta_\eta(t)= 
\delta(t).$ It follows from these arguments that the velocity-gradient experienced by the particle
should be approximated by a Gaussian matrix-valued process, constant in space and 
delta-correlated in time,  if $\alpha<1$. This approximation could break down for fixed $\eta\ll L$
if there happens to be a small advection speed $v\ll v_0.$      

Now consider the non-frozen velocity field, with covariance given by (\ref{velcov}) and (\ref{powlawcov}) 
for $D_3\neq 0.$ In this case the single-point, 2-time covariance of the velocity-gradient averaged over 
directions has the form 
\begin{eqnarray}
&&  \bar{C}_{ij,mn}(r=0,\tau) = D_1' \eta^{\alpha-2} e^{-D_3|\tau|/\eta^\beta} 
\left[(d+1)\delta_{ij}\delta_{mn}\right. \cr
&& \,\,\,\,\,\,\,\,\,\,\,\,\,\,\,\,\,\,\,\,\,\,\,\,\,\,\,\,\,\,\,\,\,\,\,\,\,\,\,\,\,\,\,\,\,\,\,\,\,\,\,\,\,\,\,\,\,\,\,\,\,\,\,\,\,\,
\left.-(\delta_{im}\delta_{jn}+\delta_{in}\delta_{jm})\right] 
\lb{velgradcovb}  \end{eqnarray} 
There is now a short correlation time $t_\eta=\eta^\beta/D_3,$ which allows us to write 
\begin{eqnarray}
&&  \langle u_{i,m}'(t)u_{j,n}'(0)\rangle = \frac{2D_1'}{D_3}\eta^{\alpha+\beta-2}  \delta_\eta(t) 
\left[(d+1)\delta_{ij}\delta_{mn}\right. \cr
&& \,\,\,\,\,\,\,\,\,\,\,\,\,\,\,\,\,\,\,\,\,\,\,\,\,\,\,\,\,\,\,\,\,\,\,\,\,\,\,\,\,\,\,\,\,\,\,\,\,\,\,\,\,\,\,\,\,\,\,\,\,\,\,\,\,\,
\left.-(\delta_{im}\delta_{jn}+\delta_{in}\delta_{jm})\right] 
\lb{velgradcovb2}  \end{eqnarray} 
with $\delta_\eta(t) = \frac{1}{t_\eta}\Delta(\frac{1}{t_\eta})$ for $\Delta(t)=  2\exp(-|t|).$
Thus, the single-point statistics of the velocity-gradient becomes temporally delta-correlated for 
vanishing $\eta.$ In addition, there is the same decorrelation effect of rapid sweeping through space 
that occurs in the frozen-field case. The latter will dominate when $\eta/v\ll \eta^\beta/D_3$ at small 
$\eta$ and when the spatial decay of correlations is fast enough, that is,  when both $\beta<1$
and $\alpha<1.$ In any case, we obtain again a model for Lagrangian velocity-gradients 
that are Gaussian, constant in space and delta-correlated in time. There is here no problem 
with small speeds $v\ll v_0,$ since the correlation time will never be larger than $\eta^\beta/D_3.$

The stability matrix has been well-studied for Gaussian velocity-gradient fields, constant in space 
and white-noise in time.  In particular, it has been shown in \cite{Bernardetal98} that the matrix 
random process $\bg(\by,\tau|0)$ is a diffusion on the group $SL(d)$ of $d\times d$ matrices 
with determinant 1. We shall use specifically the formula for the transition probability density 
$p_\tau(\bg)$ of this process starting at the identity, Eq.(7.14) in \cite{Bernardetal98} for $n=2:$
\be \int_{SL(d)} p_\tau(\bg) f(\bg\br_0) d\mu(\bg) =\int P_2(\br,\tau|\br_0,0) f(\br) \, d^dr \lb{BGK} \ee
where $\mu$ is Haar measure on $SL(d)$ and 
\be \partial_\tau P_2(\br,\tau|\br_0,0) = {\mathcal M}_2 P_2(\br,\tau|\br_0,0) \ee
for 
\begin{eqnarray} 
{\mathcal M}_2 f(\br) &=& D[(d+1)\delta_{ij}r^2-2r_ir_j]\partial_{r_i}\partial_{r_j}f(\br) \cr
                            &=& D\partial_{r_i}\partial_{r_j}\big\{[(d+1)\delta_{ij}r^2-2r_ir_j]f(\br)\big\},
\end{eqnarray} 
where the second line follows by incompressibility. This implies also that the operator is
self-adjoint. 
Since ${\mathcal M}_2f\equiv 0$ for a general linear function $f(\br)=\ba\bdot\br,$ 
and considering in (\ref{BGK}) arbitrary choices of $\ba,\br_0,$ it follows that 
\be \int_{SL(d)} \bg \, p_\tau(\bg) \,d\mu(\bg) = {\bf I}, \ee
the identity matrix. This result is due essentially to the fact that a diffusion leaves 
invariant a linear profile. Finally, we can conclude that 
\be \langle g_{jk}(\by,\tau|0)|\bzed,0;\by,\tau\rangle=\delta_{jk} 
\lb{g-delta}. \ee
The exponential growth of the individual realizations is offset by their rapid rotation
in space which leads to large cancellations in the ensemble average. Incompressibility 
was necessary to the argument.

The result (\ref{g-delta}) is only strictly known to be valid when the velocity covariance 
converges to an $\eta$-independent result as $\eta\rightarrow 0$, whereas (\ref{velgradcov2}) 
diverges as $\sim \eta^{\alpha-1}$ for $\alpha<1$ and (\ref{velgradcovb2}) diverges 
as $\sim \eta^{\alpha+\beta-2}$ for $\alpha+\beta<2.$ However, the final result (\ref{g-delta})
is independent of the amplitude of the covariance (i.e. the value of $D_1$) and thus we 
conjecture that it extends even to the present cases with diverging covariance. 
The result yields a simplified formula for the 2-particle eddy-diffusivity:  
\be K_{ij}(\br,t) =\int_{-t}^0 d\tau  \int d^dy \, S_{ij}(\br;\by,0,\tau) 
 P(\by,\tau|\bzed,0), \lb{Gauss-KLform} \ee
together with (\ref{YM102}),(\ref{gauss-v}). We shall now use this  formula to obtain  
concrete results for the eddy-diffusivity in the Gaussian ensembles whose covariances 
are given by (\ref{velcov}).

\subsection{The Frozen-in-Time Velocity Field}\lb{froz}

The simplest case to analyze is the ``frozen'' field so that
\be S_{ij}(\br;\by,0,\tau) = S_{ij}(\br;\by). \lb{Sry} \ee
Making the change of variables $u=y^2/2v_0^2\tau^2,$ 
\be \int_{-t}^0 d\tau\, P(\by,\tau|\bzed,0) = \frac{1}{\sqrt{8}\pi^{d/2}}
\frac{1}{v_0y^{d-1}} \Gamma\left(\frac{d-1}{2},\frac{y^2}{2v_0^2t^2}\right)
\ee
with the (upper) incomplete gamma function defined by $\Gamma(s,z)=\int_z^\infty du\, u^{s-1} e^{-u}.$
Since $d^dy=y^{d-1}dy\,d\Omega_y,$ with $d\Omega_y$ the element of $d$-dimensional solid angle,
we get from (\ref{Gauss-KLform}) that
\be K_{ij}(\br,t) = \frac{1}{\sqrt{2}\Gamma\left(\frac{d}{2}\right)} \int_0^\infty \frac{dy}{v_0} \,
\bar{S}_{ij}(\br;y) \Gamma\left(\frac{d-1}{2},\frac{y^2}{2v_0^2t^2}\right) \ee
where the angle-averaged structure function is defined by 
\be \bar{S}_{ij}(\br;y)=\frac{1}{S_d} \int d\Omega_y S_{ij}(\br;\by) \ee
for $S_d=2\pi^{d/2}/\Gamma\left(\frac{d}{2}\right)$ the $(d-1)$-dimensional area of the  unit 
hypersphere in $d$-dimensional space. 

When the velocity statistics are isotropic, as for the model with zero mean and covariance 
(\ref{velcov}), the eddy-diffusivity tensor can be reduced to two scalar functions $K_L,K_N$
defined by 
\be K_{ij}(\br,t) = K_L(r,t) \hat{r}_i\hat{r}_j + K_N(r,t)(\delta_{ij}- \hat{r}_i\hat{r}_j ). \ee
These two functions are related by incompressibility as $K_N=K_L+
rK_L'/(d-1)$ and it is convenient to base further analysis on $K_L.$ As is well known,
if the separation statistics are also isotropic, then the diffusion equation (\ref{Keq}) 
can be expressed entirely in terms of $K_L,$ as
\be \partial_t P(r,t)=\frac{1}{r^{d-1}}\frac{\partial}{\partial r}\left(
r^{d-1}K_L(r,t) \frac{\partial P}{\partial r}(r,t)\right). \lb{isoeq} \ee
Here the separation PDF satisfies
\be  \int_0^\infty P(r,t)\, r^{d-1}dr=1. \lb{norm} \ee
as normalization condition. 

The displacement vector $\by$ in (\ref{Sry}) breaks rotation invariance, but  the average 
over solid angle restores isotropy. We can thus decompose also 
\be \bar{S}_{ij}(\br;y) = \bar{S}_L(r;y) \hat{r}_i\hat{r}_j + 
       \bar{S}_N(r;y)(\delta_{ij}- \hat{r}_i\hat{r}_j ) \ee
into longitudinal 
and transverse contributions with respect to the separation vector $\br$. By dimensional 
analysis one can write
\be \bar{S}_L(r;y) = S_L(r) F\left(\frac{y}{r},\frac{L}{r}\right) = S_L(r) F\left(\frac{y}{r}\right), \ee
the latter for $L\gg r.$ The function $F(y/r)$ can be interpreted as the correlation 
coefficient of (longitudinal) velocity increments $\delta v_L(r)$ at points a distance $y$  
apart. For the velocity covariance (\ref{velcov}) with $D_3=0$ it is possible to derive 
a complicated, closed-form expression for the function $F(w)$ as suitable combinations of 
Gaussian hypergeometric functions of the argument $w^2.$ However, we shall not pursue 
this here. The most important property of $F,$ which follows from (\ref{2tS2}), is  
\be F(w)\sim \left\{\begin{array}{ll}
                              1 & w\ll 1 \cr 
                             (const.) w^{-(2-\alpha)} & w\gg1
                             \end{array}\right. . \ee
                             
Thus, we can write $K_L(r,t)=S_L(r)\tau(r,t)$ where
\be \tau(r,t) = \frac{1}{\sqrt{2}\Gamma\left(\frac{d}{2}\right)} \int_0^\infty \frac{dy}{v_0} \,
F\left(\frac{y}{r}\right)\Gamma\left(\frac{d-1}{2},\frac{y^2}{2v_0^2t^2}\right) \ee
is a 2-particle Lagrangian correlation time. With the substitution $y=r w,$ this becomes
\be \tau(r,t) = \frac{r}{v_0} J(x), \,\,\,\,\, x=\frac{v_0t}{r} \lb{tauJ} \ee
for 
\be J(x) =  \frac{1}{\sqrt{2}\Gamma\left(\frac{d}{2}\right)} \int_0^\infty dw \,
F(w)\Gamma\left(\frac{d-1}{2},\frac{w^2}{2x^2}\right) \lb{Jint} \ee 
For $d=3,$ $\Gamma(1,z)=e^{-z}$ and (\ref{Jint}) is a Laplace transform in the variable
$w^2.$ The most directly useful consequence of (\ref{Jint}) is the asymptotic behaviors   
\be J(x) \sim \left\{\begin{array}{ll}
                                x   & x\ll 1 \cr
                                J_\infty & x\gg 1 \cr
                              \end{array} \right. , \lb{Jx} \ee
where we have used $\int_0^\infty dv\, \Gamma\left(\frac{d-1}{2},\frac{v^2}{2}\right)=
\sqrt{2} \Gamma\left(\frac{d}{2}\right)$ and we have defined
\be J_\infty = \frac{\Gamma\left(\frac{d-1}{2}\right)}{\sqrt{2} \Gamma\left(\frac{d}{2}\right)}
\int_0^\infty dw\, F(w). \ee                               
The latter integral converges for $\alpha<1.$ 
We conclude that 
\be \tau(r,t) \sim \left\{\begin{array}{ll}
                                t & t\ll r/v_0  \cr
                                J_\infty \frac{r}{v_0} & t\gg r/v_0 \cr
                                \end{array} \right. . \lb{taufr} \ee
                                
Our result is quite similar to that obtained by \cite{ThomsonDevenish05} for the case of 
large mean velocity $\bar{\bu}$; see  their equation (8).   Some differences are that our 
eddy-diffusivity is isotropic and has the short-time behavior proportional to $t.$ However,
most importantly we see the same sweeping decorrelation effect, with the 2-particle correlation 
time at long times proportional to the sweeping time $r/v_0.$ With no such effect one 
would instead expect the correlation time to be always proportional to $t$ in the frozen-field
case. It should be emphasized that we obtain this result in the zero mean-velocity ensemble,
where  \cite{ThomsonDevenish05} have predicted different behavior. We shall compare our 
results with theirs in more detail in section \ref{predict}, where we shall also derive the quantitative
predictions of our formula for the growth of mean-square particle separations.                              

\subsection{Finite Time-Correlated Velocity Field}\lb{fin}

We now study the Gaussian model with covariance (\ref{velcov}) for $D_3\neq 0.$ 
More generally, consider any velocity field statistically homogeneous in space and stationary in time. 
Then (\ref{D-gauss-4}) together with (\ref{gauss-v}) \& (\ref{g-delta}) give
\begin{eqnarray}
&& D_{ij}(\bx'-\bx,t)=\int_0^t ds \int d^dy\, C_{ij}(\bx'-\by,t-s) \cr
&& \times \frac{1}{[2\pi v_0^2(t-s)^2]^{d/2}}
        \exp\left[-\frac{|\by-\bx-\bar{\bu}(t-s)|^2}{2v_0^2|t-s|^2}\right]
\end{eqnarray} 
Since the $\by$-integration has the form of a convolution, it is easily evaluated by a Fourier
transform:
\begin{eqnarray} 
&& \hat{D}_{ij}(\bk,t)=\int_0^t ds\, \hat{C}_{ij}(\bk,t-s)\cr
&& \,\,\,\,\,\,\,\,\,\,\,\,\,\,\,\,
\times \exp\left[i\bk\bdot\bar{\bu}(t-s)
-\frac{1}{2}v_0^2k^2(t-s)^2 \right]
\end{eqnarray} 
For the model in (\ref{velcov}) note $\hat{C}_{ij}(\bk,t)=\hat{C}_{ij}(\bk)\exp(-\gamma_k|t|)$ with
$\hat{C}_{ij}(\bk)=D_2P_{ij}(\bk)/k_L^{d+\alpha}$ and $\gamma_k=D_3k_L^\beta.$
For large $\bar{\bu},$ see \cite{ThomsonDevenish05}. Hereafter we take $\bar{\bu}=\bzed.$
Then making the change of variables $\sigma=v_0k(t-s),$ one obtains
\begin{eqnarray} 
&& \hat{D}_{ij}(\bk,t)=\frac{1}{v_0k}\int_0^{v_0kt} d\sigma\, \hat{C}_{ij}(\bk)\cr
&& \,\,\,\,\,\,\,\,\,\,\,\,\,\,\,\,\,\,\,\,\,\,\,\,\,\,\,\,\,\,\,\,\,\,\,\,\,\,\,\,\,\,\,\,\,\,\,\,
\times \exp\left[-\left(\frac{\gamma_k}{v_0k}\right)\sigma-\frac{1}{2}\sigma^2 \right]
\lb{Dhat} \end{eqnarray} 
Thus, for $t\ll 1/v_0k,$
\be \hat{D}_{ij}(\bk,t)\sim \hat{C}_{ij}(\bk) t. \lb{DCt} \ee
This implies by inverse Fourier transform that 
\be K_{ij}(\br,t) \sim S_{ij}(\br) t, \,\,\,\,\, t\ll r/v_0. \lb{KSt} \ee
On the other hand, consider fixed $t$ and large $k.$
Note that the convection time is smaller than the correlation time, or $v_0k>\gamma_k,$
for $k>k_*=(D_3/v_0)^{\frac{1}{1-\beta}}$ when $\beta<1.$ Thus, for $k\gg k_*,$
(\ref{Dhat}) gives 
\be \hat{D}_{ij}(\bk,t)\sim \frac{1}{v_0k}\hat{C}_{ij}(\bk)\cdot\sqrt{\frac{\pi}{2}}
{\rm erf}\left(\frac{v_0kt}{\sqrt{2}}\right). \lb{exact1}\ee
This formula is exact in the case of frozen turbulence ($D_3=0$) when $k_*=0.$
If furthermore $k\gg 1/v_0t,$ then 
\be \hat{D}_{ij}(\bk,t)\sim \sqrt{\frac{\pi}{2}}\cdot \frac{1}{v_0k}\hat{C}_{ij}(\bk) \ee
becomes independent of $t$ and scales as a power $k^{-(d+\alpha+1)}.$ For 
$\alpha<1,$ we thus obtain by inverse Fourier transform that for $r\ll \min\{v_0t,L_*\}$
\be K_{ij}(\br,t) \sim  \sqrt{\frac{\pi}{2}}\frac{D_1^{(\alpha+1)}}{v_0} r^{\alpha+1}
\left[(d+\alpha)\delta_{ij}-(\alpha+1)\hat{r}_i\hat{r}_j\right]. \lb{short} \ee
It follows that the essential behavior of the frozen field case carries over to the finite
time-correlated velocity with $\alpha<1$ and $\beta<1.$ Just as for the frozen velocity,
$K_L(r,t)=S_L(r)\tau(r,t)$ and the correlation time satisfies (\ref{tauJ}) and (\ref{Jx}) with 
$J_\infty =\sqrt{\frac{\pi}{2}}\frac{D_1^{(\alpha+1)}}{D_1^{(\alpha)}}$\footnote{Here 
the superscript in $D_1^{(\alpha)}$ is used to indicate the spatial scaling exponent 
$\alpha$ for which the constant in (\ref{powlawcov}) is calculated. Using eqs. (2.14)
and (2.16) in \cite{EyinkXin00} to calculate $D_1^{(\alpha)}$ gives
$$ J_\infty =\sqrt{\frac{\pi}{8}} \frac{\alpha}{\alpha+1}
\frac{\Gamma\left(\frac{1-\alpha}{2}\right)}{\Gamma\left(\frac{2-\alpha}{2}\right)}
\frac{\Gamma\left(\frac{d+\alpha+2}{2}\right)}{\Gamma\left(\frac{d+\alpha+3}{2}\right)}$$
}.

If, however, $\beta>1,$ then the behavior is quite different. Under this assumption 
$\gamma_k>v_0k $ for $k>k_*,$  so that we now make the change of variables $\sigma=
\gamma_k(t-s)$ to obtain 
\begin{eqnarray} 
&& \hat{D}_{ij}(\bk,t)=\frac{1}{\gamma_k}\int_0^{\gamma_kt} d\sigma\, \hat{C}_{ij}(\bk)\cr
&& \,\,\,\,\,\,\,\,\,\,\,\,\,\,\,\,\,\,\,\,\,\,\,\,\,\,\,\,\,\,\,\,\,\,\,\,\,\,\,\,\,\,\,\,\,\,\,\,
\times \exp\left[-\sigma-\frac{1}{2}\left(\frac{v_0k}{\gamma_k}\right)^2\sigma^2 \right]
\end{eqnarray} 
Equations (\ref{DCt}) and (\ref{KSt}) again hold, now for $t\ll 1/\gamma_k$ and $t\ll 1/\gamma_{1/r},$
respectively. On the other hand, for fixed $t$ and $k\gg k_*=1/L_*,$
\be \hat{D}_{ij}(\bk,t)\sim \frac{1}{\gamma_k}\hat{C}_{ij}(\bk)\left[
1-\exp(-\gamma_kt)\right]. \ee
If furthermore $k\gg 1/L_\beta(t),$ then 
\be \hat{D}_{ij}(\bk,t)\sim \frac{1}{\gamma_k}\hat{C}_{ij}(\bk) \ee
becomes independent of $t$ and scales as a power $k^{-(d+\alpha+\beta)}.$ When 
$\alpha+\beta<2,$ we then obtain by inverse Fourier transform that for $r\ll \min\{L_\beta(t),L_*\}$
\be K_{ij}(\br,t) \sim  \frac{D_1^{(\alpha+\beta)}}{D_3} r^{\alpha+\beta}
\left[(d+\alpha+\beta)\delta_{ij}-(\alpha+\beta+1)\hat{r}_i\hat{r}_j\right]. \ee
We can again write $K_L(r,t)=S_L(r)\tau(r,t)$ but now 
\be \tau(r,t) \sim \left\{\begin{array}{ll}
                                t & t\ll r^\beta/D_3   \cr
                                (const.) \frac{r^\beta}{D_3} & t\gg r^\beta/D_3, \, r\ll L_* \cr
                                \end{array} \right. . \ee
Thus, the sweeping decorrelation effect is absent at sufficiently small scales when
$\beta>1$ and $\alpha+\beta<2.$

\section{Consequences of Diffusion Model}\lb{predict}

In the previous section we have derived a diffusion model which, for homogeneous 
and isotropic statistics, takes the form (\ref{isoeq}).  For the Gaussian velocity ensemble 
having covariance (\ref{velcov}) with Kolmogorov scaling  exponent $\alpha=2/3,$ the 
diffusivity takes the form 
\begin{eqnarray}
K_L(r,t) &= & \frac{C_L\varepsilon^{2/3}r^{5/3}}{v_0} J\left(\frac{v_0t}{r}\right) \cr
&\sim & \left\{\begin{array}{ll}
            C_L (\varepsilon r)^{2/3} t &  \,\,\,\,\,\, t\ll r/v_0\cr
            C_L' \frac{\varepsilon^{2/3}r^{5/3}}{v_0} & \,\,\,\,\,\, t\gg r/v_0  \cr
            \end{array}\right. 
      \lb{KL} \end{eqnarray} 
both in the frozen case and in the temporally fluctuating case with $\beta=2/3$. Here $C_L$
is the Kolmogorov constant in the longitudinal velocity structure function, $S_L(r)\sim
C_L(\varepsilon r)^{2/3},$ and $C_L'=C_LJ_\infty.$ In this  section we shall attempt 
to determine the growth law for the mean-square separation
$\langle r^2(t)\rangle$ predicted by the model (\ref{isoeq}),(\ref{KL}). 

Does this model lead to the $t^{9/2}$-law of Thomson-Devenish \cite{ThomsonDevenish05}? 
To answer this question, we must briefly review the argument for the $9/2$-law. The key idea 
in \cite{ThomsonDevenish05} is that the mean-square separation pointwise in space depends
on the local value $v'$ of the fluctuating velocity. The sweeping effect occurs at points where 
$\tau_{sw}(r)=r/v^\prime$ is smaller than the intrinsic correlation time, $\tau_{int}(r,t)=
\varepsilon^{-1/3}r^{2/3}$ for finite-correlated velocity ($\beta=2/3)$ and $\tau_{int}(r,t)=t$ 
for ``frozen'' velocity.  The local correlation time is argued to be the smallest of these:  
\be \tau(r,t) =\min\{\tau_{sw}(r),\tau_{int}(r,t)\}. \lb{mintim} \ee
Hence, when $v^\prime>(\varepsilon r)^{1/3}$ (fluctuating) or $r/t$ (frozen), then the 
mean-square separation conditioned on $v'$ is affected by sweeping and shows the slow growth 
\be \langle r^2(t)\rangle_{v'} \sim \frac{\varepsilon^4t^6}{v^{\prime 6}} \lb{t6v'} \ee
but in the opposite case exhibits the faster growth 
\be \langle r^2(t)\rangle_{v'} \sim \varepsilon t^3. \lb{t3} \ee
Using these growth laws to evaluate $\tau_{sw}$ and $\tau_{int}$ in (\ref{mintim}), it 
is easily checked that the $t^6$-law holds for points with $v'>(\varepsilon t)^{1/2}$ and the 
$t^3$-law for points with $v'<(\varepsilon t)^{1/2}.$ The probability for the latter condition 
to hold is small but growing in time:
\be {\rm Prob}\Big(v'<(\varepsilon t)^{1/2}\Big)\sim \frac{(\varepsilon t)^{3/2}}{v_0^3}. 
\lb{smallvp} \ee
This formula holds for a Gaussian distribution of 3D velocities ${\bv }',$ or for any similar 
distribution $p(\bv') = (1/v_0^3)f(\bv'/v_0)$ with variance $v_0^2$ and non-vanishing density 
at the origin. The unconditional mean-square separation can then be estimated from 
(\ref{t3}) and (\ref{smallvp}) as  
\be \langle r^2\rangle \sim \varepsilon t^3 \cdot \frac{(\varepsilon t)^{3/2}}{v_0^3} 
     = \frac{ \varepsilon^{5/2} t^{9/2}}{v_0^3}. \lb{t9/2} \ee
The same result can be obtained from the $t^6$ dispersion law (\ref{t6v'}) by noting that it 
is a rapidly decreasing function of $v',$ so that the dominant contribution is obtained from 
the points with $v'\gtrsim (\varepsilon t)^{1/2}$ which also occur with probability 
$\sim (\varepsilon t)^{3/2}/v_0^3.$     

At first sight, it appears that the model (\ref{isoeq}),(\ref{KL}) may embody these ideas 
of \cite{ThomsonDevenish05}. The diffusion model implies the exact equation
\be \frac{d}{dt}\langle r^2(t) \rangle = 2\int K_T(r,t) P(r,t) r^{d-1} dr, \lb{dr2dt} \ee
where $K_T=K_L+(d-1)K_N$ is the trace of the diffusion tensor.  The average over $r$ 
in (\ref{dr2dt}) can thus play the same role as did the average over $v'$ in the argument 
of \cite{ThomsonDevenish05}. The eddy-diffusivity (\ref{KL}) is equivalent to the 
correlation time (\ref{taufr}). 
The population of particle pairs with separations $r>v_0t$ should exhibit 
a growth law $\langle r^2(t)\rangle_>\sim \varepsilon t^3$, while the pairs with 
$r<v_0t$ should exhibit  $\langle r^2(t)\rangle_<\sim \frac{\varepsilon^4 t^6}{v_0^6}.$   
It appears possible that averaging over the entire range of pair separations could give 
rise to the $9/2$-law (\ref{t9/2}) with an intermediate growth rate. 

The above reasoning is, however, essentially wrong. The diffusion model (\ref{isoeq}),(\ref{KL})
does possess a $t^3$ regime, but only in an unphysical way. To see this, note that for both
the $t^3$ and the $t^6$ growth laws the condition $r>v_0t$ is first satisfied only at such long  
times that $t>v_0^2/\varepsilon.$ Substituting the standard relation $\varepsilon \sim v_0^3/L$ 
(which follows from the assumed Kolmogorov scaling of the energy spectrum) implies 
that the $t^3$ law can be self-consisently satisfied only for times greater than a large-eddy 
turnover time, $t>L/v_0.$ In that case, $\langle r^2(t)\rangle$ exceeds $L^2$ and the particle 
pairs have left the inertial range. As we shall verify below, the model (\ref{isoeq}),(\ref{KL}) 
does indeed possess a $t^3$ range when $t>v_0^2/\varepsilon$ but this exceeds the validity
of the model, which was derived only for the range $r<L.$ In the range $r>L$ the two particles 
should instead execute independent Brownian motions with a constant diffusivity $D_T$ 
and the mean-square separation grow diffusively as $\langle r^2\rangle \sim 4 D_T t.$ 
Thus,  the $t^3$ range is an unphysical artefact of the model (\ref{isoeq}),(\ref{KL}). 

The argument for the asymptotic $t^{9/2}$ law by Thomson \& Devenish \cite{ThomsonDevenish05}
thus fails for the model (\ref{isoeq}),(\ref{KL}). Very importantly, however, we shall show below 
that our diffusion model can produce an apparent $t^{9/2}$ law over a finite range of scales 
at relatively low Reynolds numbers, for similar choices of parameters with which such
growth laws have been observed  in kinematic simulations \cite{ThomsonDevenish05,
DevenishThomson09,NicolleauNowakowski11}. In this case, the eddy-diffusivity in the 
equation (\ref{isoeq}) is not given by the formula (\ref{KL}), which is asymptotically 
valid only for $L\gg r,$ but instead directly from the expression (\ref{Dhat}), which holds 
in general. We shall thus suggest that the $9/2$ growth law observed in several kinematic
simulations is a finite-Reynolds-number effect and does not represent the asymptotic 
behavior that would be observed with very long inertial ranges.  

We argue that the true high-Reynolds-number behavior---both in our diffusion model 
and in the kinematic simulations---is essentially the same as that found by Thomson \& Devenish 
for the situation of large {\it mean} velocity $\bar{u}$ (\cite{ThomsonDevenish05}, section 3.1). 
The principal difference is that we obtain also an early-time Batchelor ballistic range 
\cite{Batchelor50, Batchelor52} with $t^2$ growth. This is followed, as argued in 
\cite{ThomsonDevenish05}, by ranges of diffusive $t^1$ growth, $t^6$ growth and 
finally by a range of $t^1$ or $t^3$ growth, depending upon whether the correct diffusivity
(\ref{Dhat}) is used for that range or whether the $r\ll L$ approximation (\ref{KL}) 
is used (inappropriately, since $r\gg L$). We have not been able to find an analytical 
solution of our model (\ref{isoeq}),(\ref{KL}) which exhibits all of the above ranges. 
In this section we shall instead argue using a simple mean-field approximation 
\be \frac{d}{dt}r^2 = 2 K_T(r,t), \,\,\,\,\, r(0)=r_0 \lb{MF} \ee
which ignores fluctuations in the random separation $r$. In the following section 
\ref{num} we shall verify our theoretical conclusions by a numerical Monte Carlo 
solution of the diffusion model. 

The Batchelor $t^2$ regime is the only one which we can derive directly from our 
diffusion model (\ref{isoeq}) without any approximation. We take as our initial 
condition for the diffusion equation the spherical delta function 
\be P_0(r)=\frac{\delta(r-r_0)}{r_0^{d-1}} \lb{P0} \ee
with all pairs initially at separation $r_0.$ If this is substituted into the exact equation
(\ref{dr2dt}), it yields
\be \left.\frac{d}{dt}\langle r^2(t) \rangle\right|_{t=0} =0, \,\,\,\, 
\left.\frac{d^2}{dt^2}\langle r^2(t) \rangle\right|_{t=0} = 2 S_T(r_0) \ee
where the trace of the short-time result (\ref{KSt}) was used,  
\be K_T(r,t) \sim 2 S_T(r) t, \lb{shortT} \ee 
for $t\ll r_0/v_0.$ Taylor series expansion then gives
\be \langle r^2(t) \rangle - r_0^2 = S_T(r_0)t^2 + O(t^3), \lb{Batch} \ee
which is the well-known result of Batchelor \cite{Batchelor50, Batchelor52}. The  
mean-field approximation (\ref{MF}) is exact in this regime, since sufficient time 
has not passed to change $r$ substantially from its initial (deterministic) 
value $r_0.$

As noted in \cite{ThomsonDevenish05}, there is an interval of times $t>r_0/v_0$ when
$r$ has still not changed substantially from its initial value $r_0.$ For $r\approx r_0$ but 
$t\gg r_0/v_0,$ the result (\ref{shortT}) is replaced with 
\be K_T(r_0,t) \sim K_T(r_0,\infty)= C'_T \frac{\varepsilon^{2/3}r_0^{5/3}}{v_0} \ee
where $C'_T=\frac{14}{3}C_L'$ as a consequence of incompressibility. The growth law 
then becomes diffusive
\be \langle r^2(t) \rangle - r_0^2 \sim 2K_T(r_0,\infty)t, \lb{Kraich} \ee 
this period lasting until the ``takeoff time'' $t_{to}$ when $K_T(r_0,\infty)t_{to}\sim r_0^2,$ or
\be t_{to} \sim \frac{v_0r_0^{1/3}}{\varepsilon^{2/3}}. \lb{to} \ee
See \cite{ThomsonDevenish05}. Together with the previous Batchelor regime, this diffusive 
range is obtained from the mean-field model (\ref{MF}) simplified to $dr^2/dt=2K_T(r_0,t).$
It is interesting that the diffusive behavior (\ref{Kraich}) at early times is the analogue in the 
Kraichnan white-noise advection model \cite{Kraichnan68,Falkovichetal01} of the Batchelor 
ballistic range (e.g. see \cite{Eyink11}, section II.B). This is not an accident. The large-scale
sweeping of particle pairs through stationary eddies produces an effective small correlation 
time $r_0/v_0$ which makes the velocity field appear to be temporally white-noise for times 
$t\gg r_0/v_0.$ This is closely connected with previous attempts to simulate the Kraichnan 
white-noise ensemble by sweeping fixed large-scale velocity fields rapidly across the 
computational domain \cite{ChenKraichnan98,FrischWirth96}.

For times greater than the ``takeoff time'' $t_{to}$ but smaller than the ``end-of-sweeping time'' 
$t_{es}=v_0^2/\varepsilon,$ one must solve the mean-field equation (\ref{MF}) with 
\be dr^2/dt = 2C_T' \frac{\varepsilon^{2/3}r^{5/3}}{v_0}, \label{MF1} \ee 
which leads to the $t^6$-law (\ref{t6}). Instead for $t>t_{es}$ one must solve 
\be dr^2/dt =2C_T (\varepsilon r)^{2/3} t, \,\,\,\,\,\, C_T=\frac{11}{3}C_L  \label{MF2} \ee
at least for the model (\ref{KL}). As previously discussed, this leads to the Richardson $t^3$-law
but in an unphysical way, since $r>L$ lies outside the validity of the model (\ref{KL}). For $t>t_{es}$ 
and $r>L$ in reality $K_T(r,t)\sim 2 D_T,$ where $D_T$ is the 1-particle diffusivity of Taylor
\cite{Taylor21}. Thus, one must solve
\be dr^2/dt = 4 D_T \ee
which yields the very long-time diffusive range. 

Our picture of particle dispersion in the zero-mean synthetic turbulence ensembles is thus 
very close to that in the large mean-velocity ensembles. This is in contrast to Thomson \& 
Devenish \cite{ThomsonDevenish05}, who argue for a distinct behavior of particle 
dispersion in the two cases. To understand better why we reach a different conclusion, 
it is useful to rederive our results for the eddy-diffusivity in a slightly different way. For 
convenience we consider only the case of frozen velocity fields. Taking the longitudinal 
component of the formula (\ref{Gauss-KLform}) yields
\be K_L(r,t) = \int^0_{-t} d\tau \int d^dy \, S_L(\br;\by) P(\by,\tau|\bzed,0). \ee 
As discussed in section \ref{struc1pdf} the factor $P(\by,\tau|\bzed,0)$ arises as the 
density of the Gaussian large-scale velocity $\bv=\by/\tau.$ Changing to this variable
in the above integral yields
\be K_L(r,t) = \int K_L(r,t|v) \exp\left(-\frac{v^2}{2v_0^2}\right) \frac{d^dv}{(2\pi v_0^2)^{d/2}} 
\lb{KLavrg} \ee
with 
\be K_L(r,t|v) = S_L(r)\tau(r,t|v) \ee
and 
\be \tau(r,t|v)= \int^0_{-t} d\tau \, F\left(\frac{v|\tau|}{r}\right) \sim \left\{
       \begin{array}{ll}
        t & t\ll r/v \cr
        I_\infty \frac{r}{v} & t\gg r/v \cr
        \end{array} \right. \ee
for $I_\infty=\int_0^\infty dw \, F(w).$ Since $F(v|\tau|/r)$ is the correlation coefficient of 
increments $\delta u(r)$ at distance $v|\tau|$ apart, $K_L(r,t|v)$ and $\tau(r,t|v)$ can be interpreted 
as pair diffusivity and correlation time for given large-scale velocity magnitude $v$. It is easy 
to average these quantities over $v$ and recover the previous results for $K_L(r,t)$ and
$\tau(r,t),$ in particular formula (\ref{taufr}), and our predictions in this section for $\langle r^2(t)\rangle.$ 
It was already observed in \cite{ThomsonDevenish05} (section 3.2, p. 292) that averaging 
the pair-diffusivity over the large-scale sweeping velocity would lead to the $t^6$-law 
also for the zero-mean velocity ensembles. Thomson \& Devenish argued, however, 
that correct results should be obtained by averaging $\langle r^2(t)\rangle_{v}$ rather than 
by averaging $K_L(r,t|v).$ We find that the opposite is true. The exact integration-by-parts 
identity for Gaussian velocity fields leads to our formula (\ref{KLavrg}) in which the effective
diffusivity is indeed averaged over large-scale sweeping velocity.  

\section{Numerical Simulations}\lb{num}

We now present numerical results for the diffusion models derived in the previous 
sections, both to confirm our theoretical predictions of their behavior and to obtain 
new conclusions where no analytical results are available. 

\subsection{Methods and Tests}\lb{NumMeth}

As in \cite{ThomsonDevenish05}, we shall solve the diffusion equation (\ref{Keq2}) using 
a Monte Carlo method for the equivalent (Ito) stochastic differential equation
\be
dr_i=b_{ij}(\br,t) dW_j(t),\,\,\,\, i,j=1,...,d   \lb{stochas}
\ee
where Einstein summation convention is used, $W_j(t)$ is a vector Wiener process and 
$2K_{ij}=b_{ik}b_{jk},$ with lower-triangular
square-root $b_{ij}$ calculated by Cholesky decomposition.  We can integrate the stochastic 
equations (\ref{stochas}) using the standard Euler-Maruyama scheme:
\begin{eqnarray}
r_i(t_k) &=&r_i(t_{k-1})+ b_{ij}(\br,t_{k-1}) \sqrt{\Delta t} \, N_{k,j} \,\,\,\, i,j=1,...,d  \cr
t_k &=& t_{k-1}+ \Delta t  \lb{EMscheme}
\end{eqnarray}
where $N_{k,j}$ for $j=1,..,d,$ $k=1,2,3,...$ is an independent, identically distributed 
sequence of standard normal random variables. The normal random variables are 
obtained from uniform pseudorandom numbers generated by the Mersenne 
Twister algorithm \cite{MatsumotoNishimura98} which are then transformed to normal 
by the Box-Muller method \cite{BoxMuller58}.  

Unfortunately, the ranges of time that we must cover are so large that it is completely 
impossible for us to use a constant timestep $\Delta t.$ Instead we use an adaptive 
scheme similar to that of  \cite{ThomsonDevenish05}.  The stepsize is determined over 
geometric intervals $T(m)<t<T(m+1)$ with 
\be
T(m)=A\exp(Bm)    \mbox{\hspace{5ex} for } m=1,2,...,M.
\ee
The constants $A$, $B$ and $M$ are chosen for an initial particle separation $r_0$ as
\begin{eqnarray}
M&=&-\frac{250}{\ln(10)}\ln r_0+2251\\
B&=&\frac{\ln(10^9/r_0)}{M-2}\\
A&=&10^{-5}r_0\exp(-B)\\
\end{eqnarray}
so that $T(1)\ll\tau_{sw}$, $T(L)\gg t_{es}$ 
and $\Delta T\equiv T(m+1)-T(m) \ll t$. In each such interval we take 
\be
\Delta t = C_{\Delta}\mbox{min} \left(\frac{r^2}{K_{T}(r,t)},\Delta T\right)
\ee
where $K_{T}$ is the trace of $K_{ij}$. A large number $S$ of independent samples of the process
(\ref{stochas}) are generated with initial separations $\mathbf{r}(t=0)=\mathbf{r}_0$ uniformly 
distributed over a sphere of radius $|\mathbf{r}_0|$, and statistics obtained by averaging over realizations. 
Most of the results presented below used $S=10^4$.

There is considerable debate in the literature, however, whether such adaptive time-stepping schemes lead to 
converged, unbiased results for the statistics \cite{ThomsonDevenish05, Osborneetal06, DevenishThomson09, 
NicolleauNowakowski11}. To test our numerical methods, we found it useful to consider somewhat simpler 
diffusion models where exact analytical results are available for comparison. The models with a 
power-law diffusivity 
\be K_L(r)=D r^\zeta, \,\,\,\,0<\zeta<2 \lb{pow-diff} \ee 
have been very well studied. It has been shown that the long-time evolution is self-similar, with a 
dispersion law 
\be \langle r^2(t)\rangle \sim g (Dt)^{2/\gamma}, \,\,\,\,
      g = \frac{\gamma^{4/\gamma}\Gamma\left(\frac{d+2}{\gamma}\right)}{\Gamma\left(\frac{d}{\gamma}\right)} 
      \lb{Rich-disp} \ee
and a stretched-exponential PDF
\be
P(r,t)= \frac{1}{\langle r^2(t)\rangle^{d/2}}\exp\left[-\alpha \Big(\frac{r}{\langle r^2(t)\rangle^{1/2}}
\Big)^\gamma+\beta\right]
\lb{SS-PDF} \ee
where $\gamma=2-\zeta,$
\be \alpha=\left[\frac{\Gamma((d+2)/\gamma)}{\Gamma(d/\gamma)}\right]^{\gamma/ 2}, \lb{alpha} \ee
\be \beta=\ln\left[\frac{\gamma (\Gamma((d+2)/\gamma))^{d/2}}{(\Gamma(d/ \gamma))^{(d+2)/2}}\right], 
\lb{beta} \ee
with the normalization condition $\int_0^{\infty}r^{d-1}P(r,t)dr=1.$ See  \cite{HentschelProcaccia84}, 
eqs.(3.14),(3.22) and the general, self-similar solutions found in \cite{EyinkXin00} for the case $\ell=0$
 \footnote{The equations of \cite{HentschelProcaccia84} are unfortunately marred 
by several misprints}. Incidentally, note that the mean-field equation (\ref{MF}) leads to power-law growth 
with the same exponent $2/\gamma$ as in (\ref{Rich-disp}) but with a different prefactor $g^{MF}=
(\gamma(d+\zeta))^{2/\gamma}$ than $g$. It is not hard to show that $g^{MF}>g$, with 
$g^{MF}\rightarrow g$ as $d\rightarrow\infty$ from Stirling's approximation. 

Notice that the inertial-range model (\ref{KL}) reduces to the time-independent 
diffusivity 
\be K_L(r,\infty)=C_L' \varepsilon^{2/3}r^{5/3}/v_0, \lb{PLdiff} \ee 
as long as $r\ll v_0t.$ This is a special
case of the power-law diffusivity (\ref{pow-diff}) with $\zeta=5/3,$ or $\gamma=1/3,$ so that the  
mean-square separation grows as $t^6.$ This case is thus most suitable to test our numerical 
methods. For the purposes of comparison in the next section with the more complex model (\ref{KL}),
we take $D=C_L' \varepsilon^{2/3}/v_0$ with 
$C_L'=1.262$ and $d=3$ so that  
\be  \langle r^2(t)\rangle \sim g_6^{PD} \frac{\varepsilon^4 t^6}{v_0^6} \lb{t6law} \ee
with the power-law diffusion model predicting $g_6^{PD}\doteq 15.968.$ This model also 
has the self-similar PDF of form (\ref{SS-PDF}) with $d=3$, $\gamma=\frac{1}{3},$ so 
$\alpha\doteq 11.3714,$ $\beta\doteq 10.1767.$ 

\begin{figure}[!h] 
\begin{center}  
\includegraphics[width=\linewidth,,height=9cm]{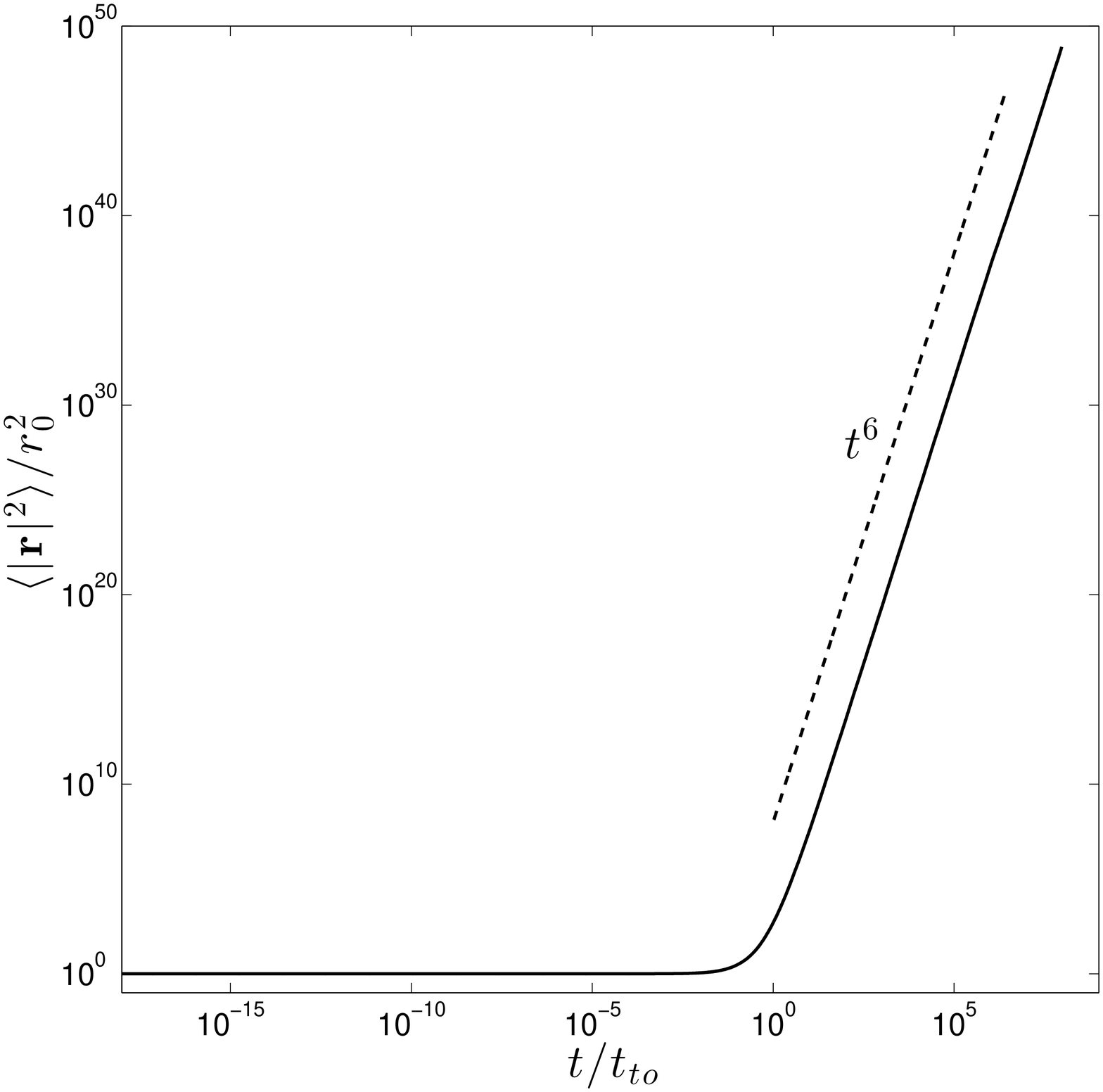}
\caption{Monte Carlo results for $\langle|\mathbf{r}(t)|^2 \rangle$ in the power-law 
diffusion model (\ref{PLdiff}) with $S=10^4$ and $C_\Delta=1$.}
\label{figure1}
\end{center} 
\end{figure}

\begin{figure}[!h] 
\begin{center}  
\includegraphics[width=\linewidth,,height=6cm]{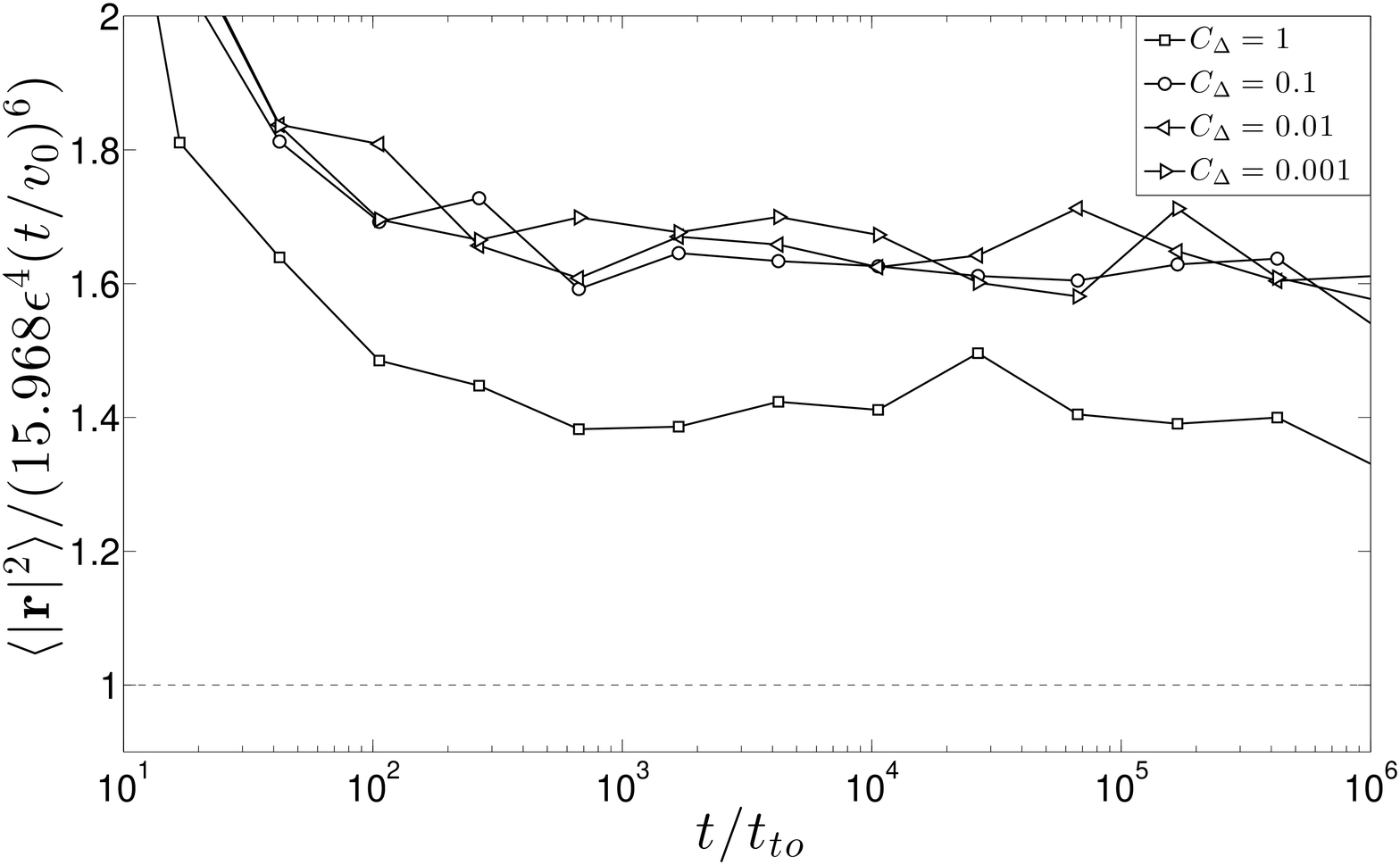}
\caption{Monte Carlo results for $\langle|\mathbf{r}(t)|^2 \rangle$ in the power-law 
diffusion model (\ref{PLdiff}) with $S=10^4$ and various $C_\Delta$, compensated by the 
analytical result (\ref{t6law}).}
\label{figure2}
\end{center} 
\end{figure}

\begin{figure}[!h] 
\begin{center}  
\includegraphics[width=\linewidth,,height=9cm]{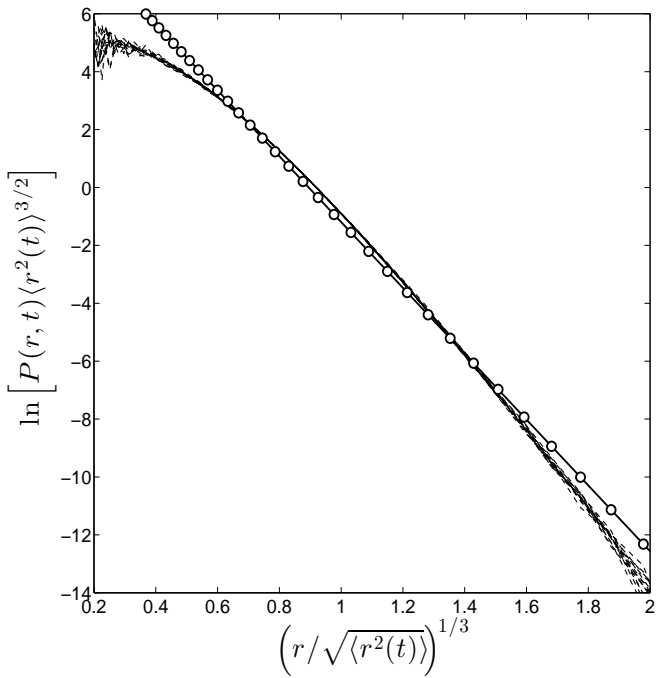}
\caption{Logarithm of the rescaled PDF of pair-separations at $14$ different  times 
in the $t^6$ range, for the power-law diffusion model (\ref{PLdiff}). Monte Carlo 
results for $S=10^5$ and $C_\Delta=1$. The straight line marked with circles
($\circ$) is the analytical result (\ref{SS-PDF}) for $d=3$ and $\gamma=1/3$.}
\label{figure3}
\end{center} 
\end{figure}

We now employ the numerical scheme discussed earlier to see which of these exact 
results we can successfully reproduce. As we see in Fig.~\ref{figure1}, long ranges 
of perfect $t^6$ power-laws can be obtained in log-log plots.
On the other, Fig.~\ref{figure2} is a semilog plot of the dispersion 
compensated by the analytical result (\ref{t6law}). It shows that the prefactor 
is poorly calculated by our Monte Carlo, which gives $g_6^{MC}\doteq 26.10$ and 
does not appear to converge to the analytical result $g_6^{PD}$ as $C_\Delta$ is decreased.  
Finally, Fig.~\ref{figure3} shows the logarithm of the PDF of pair separations $r$ 
plotted versus $r^{1/3}$ at $14$ different times in the long $t^6$-range. Self-similarity is well-confirmed 
by collapse of rescaled curves for different times, but the analytical result (\ref{SS-PDF}) 
is not very accurately reproduced.

Our conclusion from these exercises is that the adaptive time-stepping scheme should be adequate 
for exponents of dispersion power-laws, but not for prefactors or PDFs. Since the primary issue 
in this work is the exponents, we shall employ the adaptive schemes when necessary to cover 
extensive ranges where constant time-steps are unfeasible. As additional checks on our numerical 
results for exponents from adaptive schemes, we test for convergence using constants $C_{\Delta}$ 
ranging from $1$ to $10^{-6}.$ We also compare our Monte Carlo results for the diffusion 
equation with a separate numerical solution of  the mean-field equation (\ref{MF}), 
integrated with a Fortran 90 implementation of the Watt and Shampine RKF45 ODE
solver \cite{Burkhardt,Shampineetal76}. This standard ODE integration method is also 
adaptive, but with variable time-step determined by preselected error tolerances. We therefore can 
have confidence that the numerical results for the mean-field theory are well converged.  

\subsection{The Inertial-Range Model}\lb{IRnum}

\begin{figure}[!t] 
\begin{center}  
\includegraphics[width=\linewidth,,height=9cm]{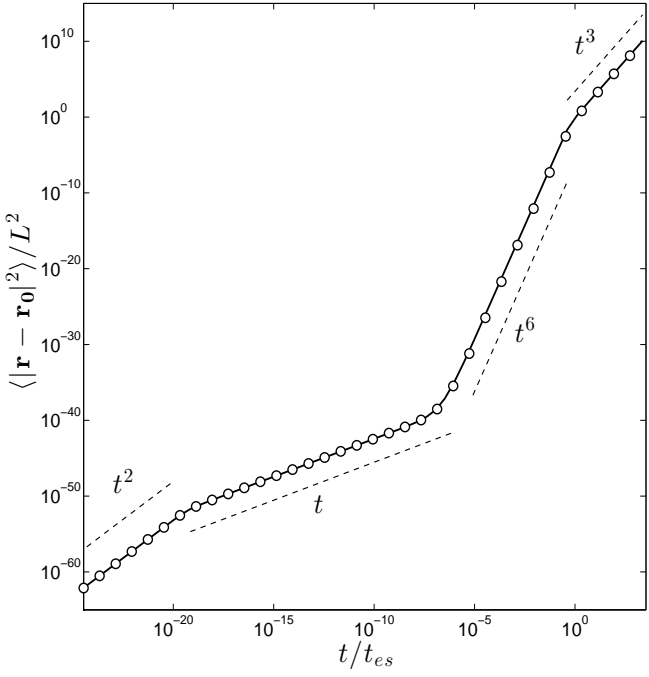}
\caption{Numerical results for $\langle|\br(t)-\br_0|^2\rangle$ in the inertial-range model 
(\ref{KL}): Monte Carlo solution of the diffusion equation ($\circ$) 
with $C_\Delta=1,S=10^4$ and mean-field approximation ({\bf --}). 
}
\label{figure4}
\end{center} 
\end{figure}

We consider first the model (\ref{KL}) obtained for Kolmogorov scaling exponents in the limit 
$L\gg 1$ and thus physically applicable only for separations $r$ in the inertial range of scales.  
This diffusion model applies for both the frozen velocity case and the finite-time correlated case
(since $\beta=2/3<1$).  For the purpose of simplifying the numerical work, we opted not to use 
the exact scaling function $J(x)$ given by integral (\ref{Jint}), which in three dimensions yields 
a complicated expression in terms of generalized hypergeometric functions. Instead, we built
a function with the same asymptotic behaviors (\ref{Jx}) as the true $J(x).$ We took 
\begin{eqnarray}
J(x)= J_\infty\mbox{erf}(\lambda x) &= & \left\{\begin{array}{ll}
            x  &  \,\,\,\,\,\, x\ll 1\cr
            J_\infty & \,\,\,\,\,\, x\gg 1  \cr
            \end{array}\right. 
\end{eqnarray} 
with $\lambda = \frac{\sqrt{\pi}}{2J_\infty}$ and $J_\infty=\sqrt{\frac{\pi}{2}}\frac{D_1^{(5/3)}}{D_1^{(2/3)}}
\doteq  0.6396$. Our expectation was that only these general features should be sufficient 
to observe the scaling regimes predicted in the previous section. This idea was borne out by 
the numerical results. In Fig. \ref{figure4} we plot $\langle|\mathbf{r}(t)-\mathbf{r}_0|^2 \rangle$ 
for the inertial-range diffusion model with $r_0=10^{-20}$. On the same graph we plot for comparison
the numerical solution $r^2(t)-r_0^2$ of the mean-field equation (\ref{MF}). The two agree very 
well, and clearly exhibit the four predicted regimes with power-laws $\propto t^2,t^1,t^6$ and 
$t^3,$ successively.  A convergence analysis of our adaptive scheme for these results is presented
in Appendix A. 

\begin{figure}[!p] 
\begin{center}  
\includegraphics[width=\linewidth,height = 5.2cm]{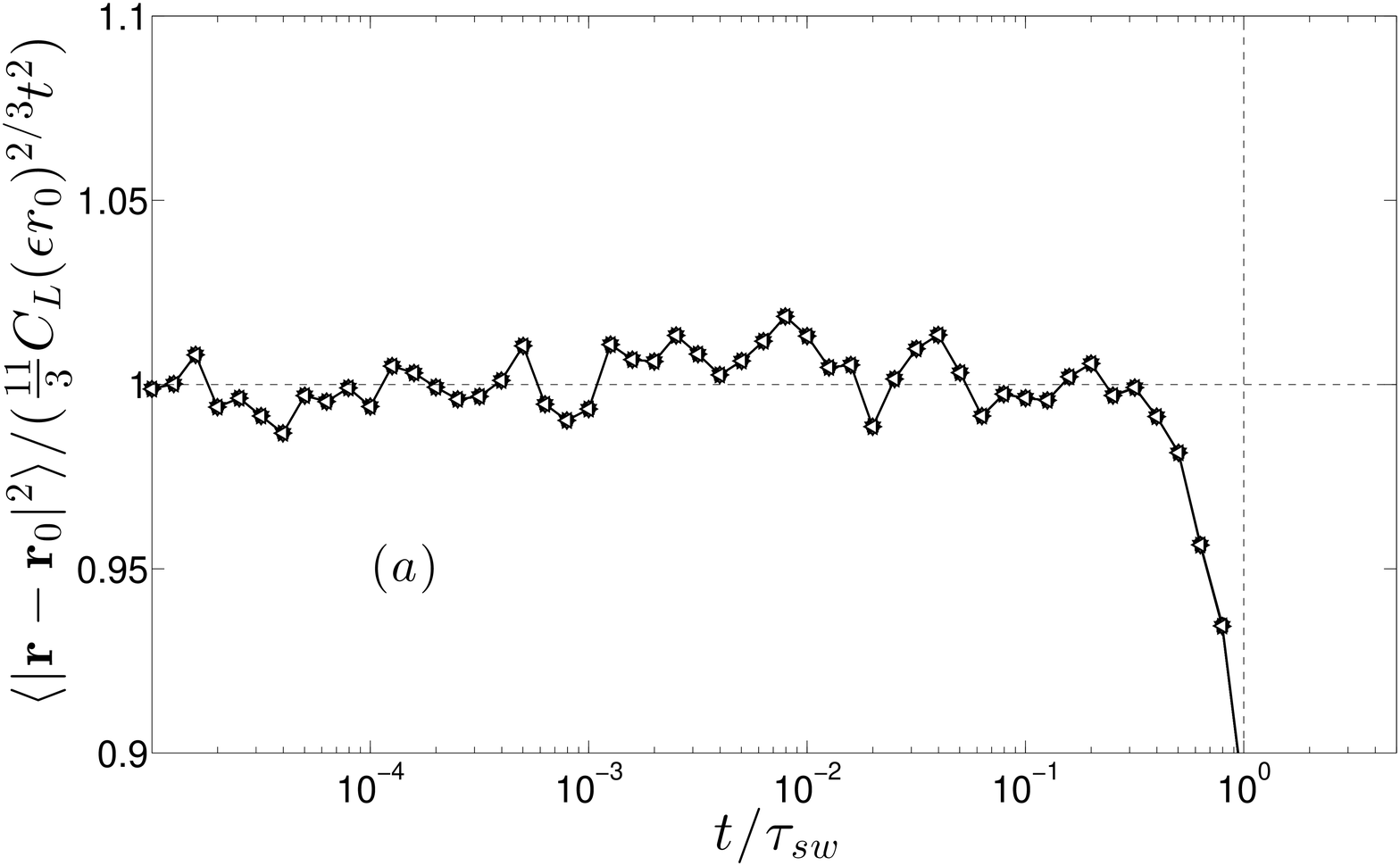}
\includegraphics[width=\linewidth,height = 5.2cm]{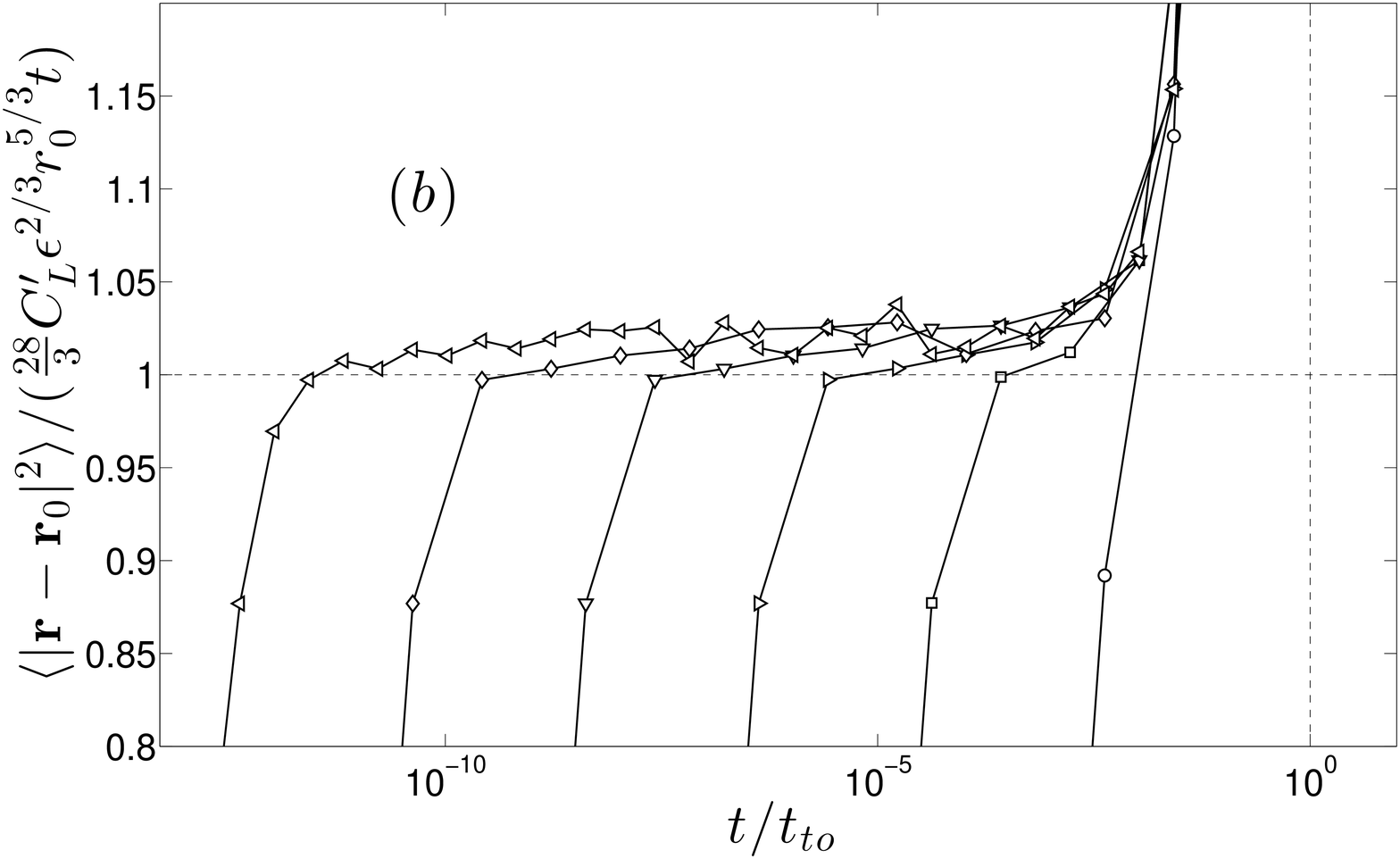}
\includegraphics[width=\linewidth,height = 5.2cm]{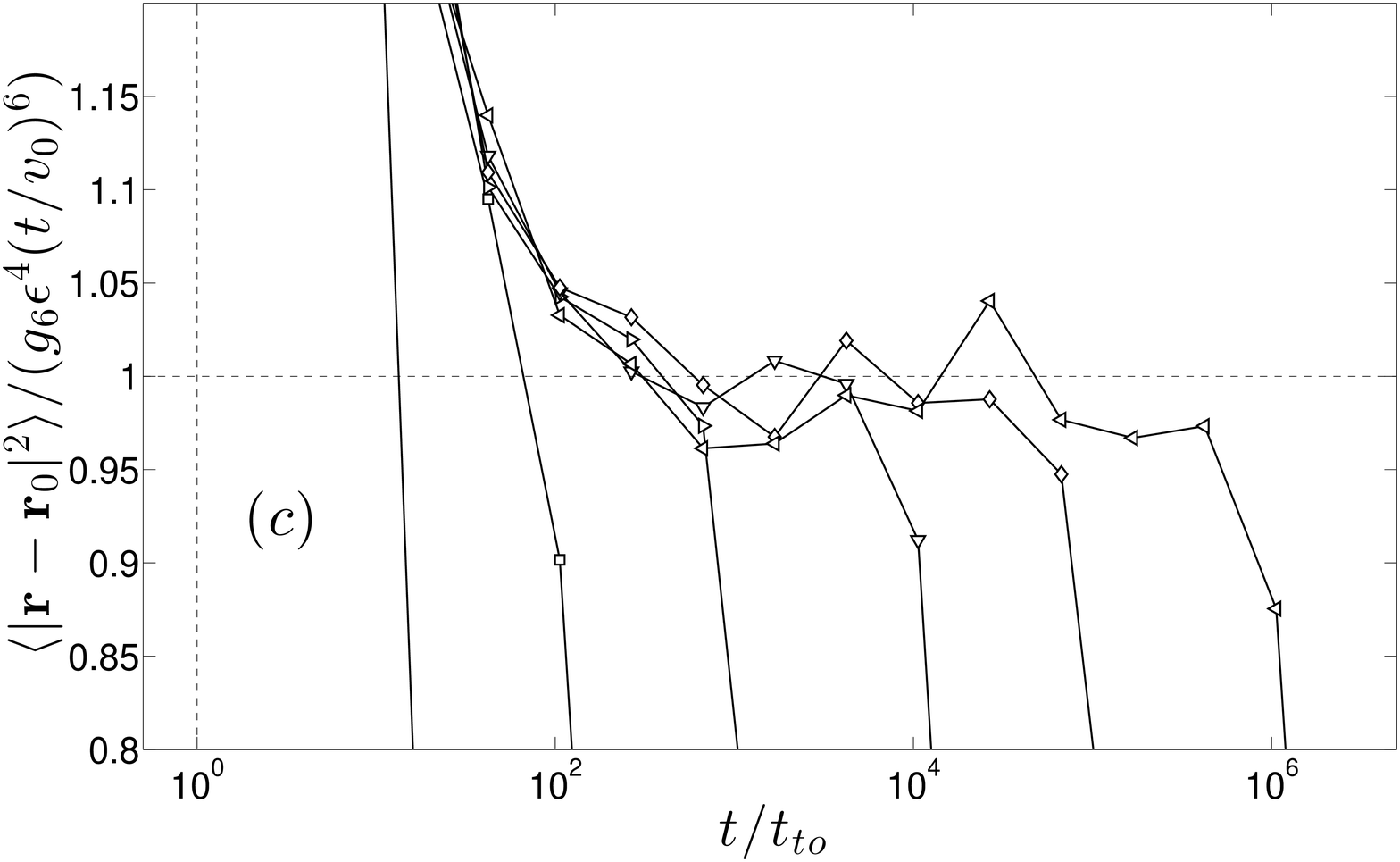}
\includegraphics[width=\linewidth,height = 5.2cm]{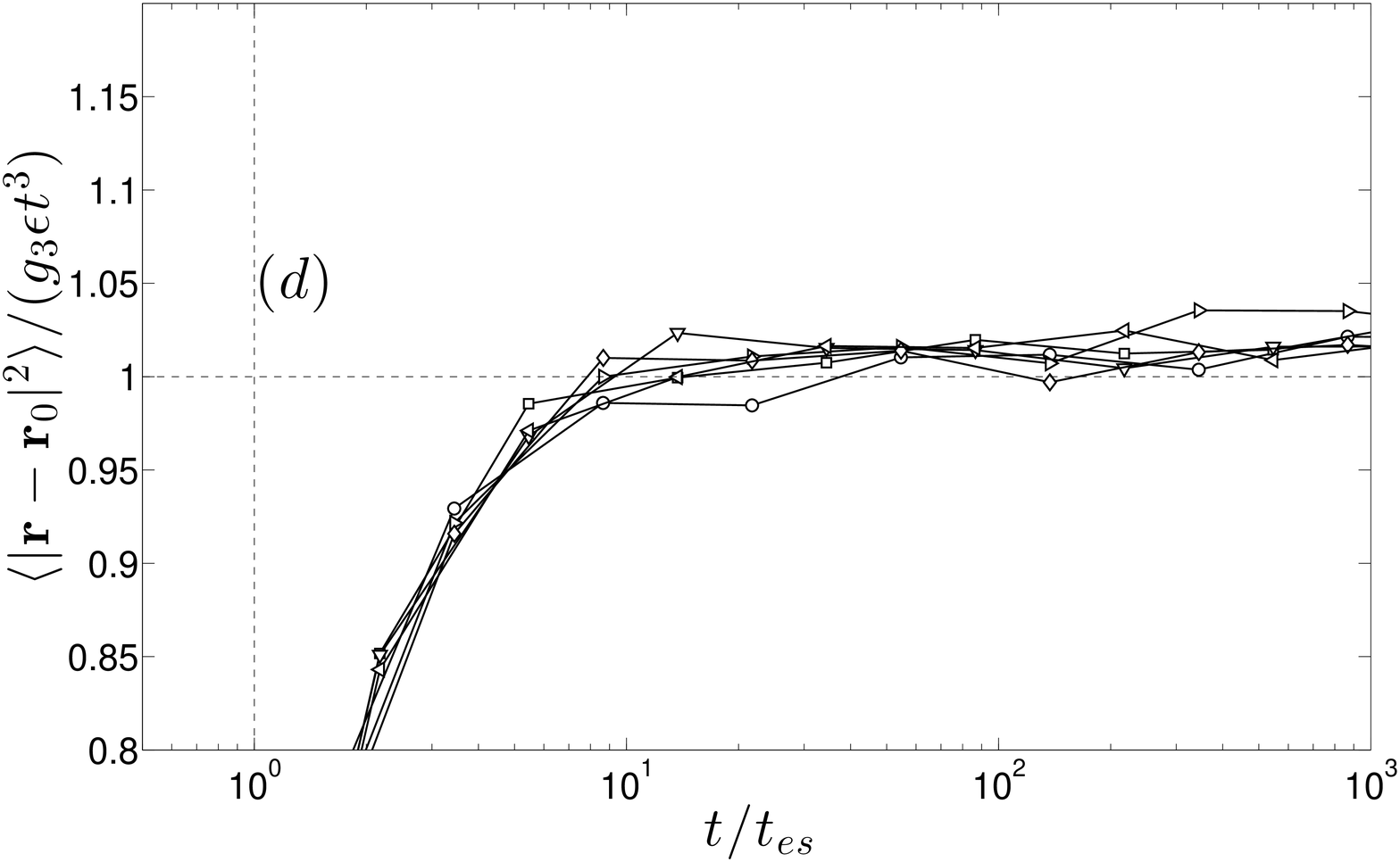}
\caption{Monte Carlo results for $\langle|\mathbf{r}(t)-\mathbf{r}_0|^2 \rangle,$ $C_\Delta=1,$ $S=10^4.$
Each panel shows the same curves with different scalings. (a) Batchelor regime. (b) Kraichnan regime. 
(c) $t^6$ regime. (d) Richardson regime. The initial separations are $r_0/L=10^{-5} (\circ)$,
$10^{-8} (\mathsmaller{\square})$ , $10^{-11} (\triangleright)$, $10^{-14} (\triangledown)$, $10^{-17} (\diamond)$,
$10^{-20} (\triangleleft).$} 
\label{figure5}
\end{center} 
\end{figure}

To further test the theoretical predictions, we investigate the crossover times between the different 
regimes and the prefactors of the scaling laws.  For example, in  Fig. \ref{figure5}(a) we show for 
various values of $r_0/L$ the quantity $\langle|\mathbf{r}(t)-\mathbf{r}_0|^2 \rangle$ 
compensated by the Batchelor-range prediction $\frac{11}{3}C_L
(\epsilon r_0)^{2/3}t^2$ plotted versus the time $t/\tau_{sw}$ rescaled with the sweeping time 
$\tau_{sw}=r_0/v_0.$ The Batchelor prediction fits the Monte Carlo data to within $0.13\%$ 
relative error and the end of  this regime is very close to $t/\tau_{sw}=1.$ We similarly show in Fig. \ref{figure5}(b) 
for the same choices of $r_0/L$ the mean-square separation $\langle|\mathbf{r}(t)-\mathbf{r}_0|^2 \rangle$ 
compensated by the Kraichnan-like ``diffusive-range'' prediction $\frac{14}{3}C_L^{\prime}\epsilon^{2/3}r_0^{5/3}t$ 
plotted versus $t/t_{to}$ with the ``takeoff time'' $t_{to}$ given by equation (\ref{to}). The diffusive-range
prediction is verified with a $1.8\%$ error and the end of this regime quite convincing scales 
as $\sim 10^{-2}t_{to}$.  In Fig. \ref{figure5}(c) we show the corresponding plot of mean-square 
separation compensated by $\varepsilon^4t^6/v_0^6$ versus $t/t_{to}.$ We see that a $t^6$ range 
begins at time $\sim 10^2t_{to}$ and extends to the end-of-sweeping time $t_{es}=v_0^2/\varepsilon$ with 
a prefactor $g_6^{MC}\simeq 22.96$ of the $t^6$-law. This Monte Carlo value lies between the mean-field prediction 
$g^{MF}_6 \doteq 55.61$ and the exact power-law diffusion model prediction $g_6^{PD}\doteq 15.97,$ 
but is quite close to the Monte Carlo result $g_6^{PMC}\doteq26.10$ for the latter model. This suggests that the adaptive 
time-integration scheme underestimates the effects of diffusion, whereas the true value for the 
inertial-range model (\ref{KL}) is probably $g_6 \doteq 15.97,$ the same as the power-law model (\ref{PLdiff}). 
It is interesting that the transition  between the $t^1$ and $t^6$ scaling ranges is quite broad, 
covering about four decades. We show finally in Fig. \ref{figure5}(d) the mean-square separation 
compensated by the Richardson prediction $\varepsilon t^3$ plotted versus $t/t_{es}.$  For $t>t_{es}$ 
there is a clear $t^3$ regime with Richardson constant $g_3^{MC}\simeq 8.97$. This constant again lies 
between mean-field predictions and exact results for a self-similar solution and should not be regarded
as accurate. Of course, as emphasized earlier, this entire regime of the inertial-range diffusion model is 
unphysical and will not be observed in KS model simulations. 

\subsection{Comparison with KS Models}\lb{KScomp}

Our derivation of diffusion model approximations was sufficiently general that we can consider 
cases of more direct relevance for KS simulations, with any energy spectrum and without the 
approximation of large $L.$ Using the formula (\ref{exact1}), which is exact for frozen turbulence, 
 one obtains by inverse Fourier transform in 3D that 
\be D_{ij}(\br,t) = \sqrt{\frac{\pi}{2}}\int \frac{\widehat{C}_{ij}(\bk)}{v_0k} {\rm erf}\left(\frac{v_0kt}{\sqrt{2}}\right)
      e^{i\bk\bdot\br}\, d^3k. \lb{KSD} \ee
It is convenient to assume statistical isotropy, so that  
\be
\widehat{C}_{ij}(\mathbf{k})=\frac{E(k)}{4\pi k^2}P_{ij}(\mathbf{k}), 
\ee
where $P_{ij}(\bk)$ is the projection operator onto the subspace orthogonal to $\bk.$  The trace 
of the diffusivity tensor becomes
\be D_T(r,t) = \frac{\sqrt{2\pi}}{v_0} \int_0^\infty \frac{dk}{k} E(k) {\rm erf}\left(\frac{v_0kt}{\sqrt{2}}\right)
     \frac{\sin(kr)}{kr} \lb{DTrt} \ee
and $D_L(r,t)$ can be recovered from 
$$ D_L(r,t) =\frac{1}{r^d}\int_0^r D_T(\rho,t)\,\rho^{d-1}d\rho. $$
Finally, the diffusivity that appears in equation (\ref{isoeq}) is 
$$ K_L(r,t) = 2(D_L(0,t)-D_L(r,t)). $$

To apply these results to the KS models
\cite{ThomsonDevenish05, Osborneetal06, DevenishThomson09, NicolleauNowakowski11},
let us recall that those models have a discrete set of wavenumbers distributed as 
\be
k_n=k_1\left(\frac{k_N}{k_1}\right)^{\frac{n-1}{N-1}}, 
\ee
for $n=1,...,N$ where $k_1=2\pi/L$, $k_N=2\pi/\eta$ and $\eta$ is the analogue of the Kolmogorov 
dissipation length. The energy spectrum generally adopted in these models  is
\be
E(k)=C_K \varepsilon^{2/3}\sum_{n=1}^Nk_n^{-5/3}\delta(k-k_n)\Delta k_n
\lb{KSspec} \ee
where $\Delta k_n=(k_{n+1}-k_{n-1})/2$ and $C_K=1.5$ is the Kolmogorov constant,   
so that $C_L\doteq 1.973$ \footnote{The standard formula $C_L=\frac{2\pi}{(3+\alpha)\Gamma(2+\alpha)\sin(\pi\alpha/2)}C_K$ 
relates the constants $C_L,C_K$ for a $k^{-(1+\alpha)}$ power-law spectrum;  e.g. see \cite{YaglomMonin75}, eq.(13.100).
This leads to $C_L\doteq 1.9727$ and $C_L'=C_LJ_\infty\doteq 1.262,$
the choice of the previous two subsections.}.
Here $\varepsilon$ is a constant with dimensions of energy dissipation per mass chosen 
to prescribe values of the rms velocity:
\be
v_0=\sqrt{\frac{2}{3}\int_{k_1}^{k_N}E(k)dk}.
\ee
The formula (\ref{DTrt}) with the KS spectrum (\ref{KSspec}) yields
\be
D_T(r,t)=\frac{C_K\epsilon^{2/3}}{v_0r }\sqrt{2\pi}\sum_{n=1}^N 
\frac{\mbox{erf}\left(\frac{v_0k_nt}{\sqrt{2}}\right)}{k_n^{11/3}}\sin\left(k_nr\right)\Delta k_n
\lb{KSDT} \ee
The assumption of isotropy in this formula is only approximately valid for KS simulations. It would 
be possible to use the general result (\ref{KSD}), without assuming isotropy, which would lead to a 
discrete sum over wavevectors rather than wavevector magnitudes. However, this would make numerical
implementation a bit more difficult, without essentially different physics. 
  
We now present simulation results for diffusion models based on Gaussian velocity fields with the spectra of 
KS models, or, to be brief,  ``KS diffusion models'' .  The same Monte Carlo method was employed as for the 
inertial-range model. In all of our  numerical studies we take $v_0=L=1$. We tried various values for the 
number of modes $N$ and we found that the numerical results on dispersion laws in log-log plots for 
$N\gtrsim100$ are not significantly different (see Appendix B). All of our presented results are for $N=500,$ a 
comparable number to that in the KS studies \cite{ThomsonDevenish05, Osborneetal06, DevenishThomson09, 
NicolleauNowakowski11}. We have also followed the practice in the KS literature of choosing the smallest 
length-scale $\eta=r_0/10,$ for initial separation $r_0.$

\begin{figure}[!p] 
\begin{center}  
\includegraphics[width=\linewidth,,height=9cm]{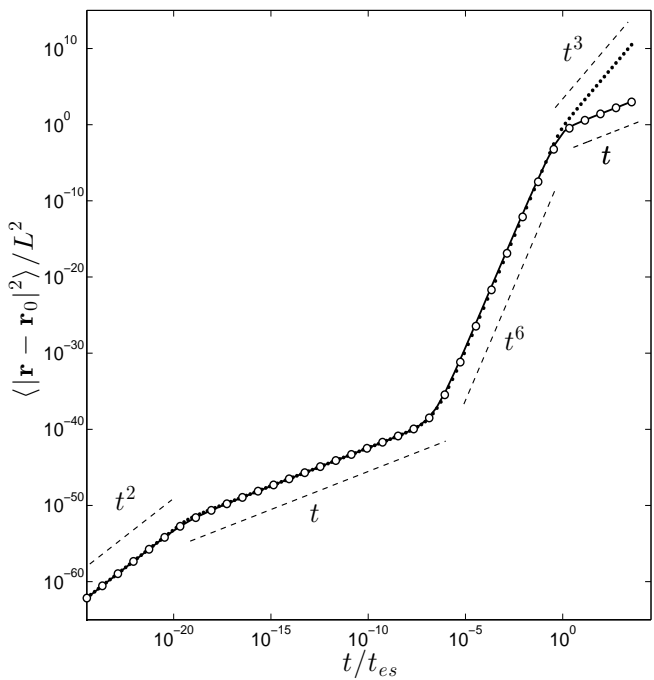}
\caption{Numerical results for $\langle|\br(t)-\br_0|^2\rangle$ in the KS diffusion model (\ref{KSDT}),
$k_N/k_1=10^{21}$: Monte Carlo solution of the diffusion equation ($\circ$) with $C_\Delta=1,
S=10^4$ and mean-field ({\bf --}). Also MC results (\boldmath{$\cdots$}) for inertial-range model (\ref{KL}).} 
\label{figure6} 
\end{center} 
\end{figure}

Our first set of numerical experiments investigated whether these more realistic models would exhibit
the power-law scaling ranges  predicted in section \ref{predict}, with a $k_N/k_1$ sufficiently large. 
In Fig. \ref{figure6} we plot the numerical results for the mean-square separation 
$\langle |\br(t)-\br_0|^2\rangle$ obtained from the KS diffusion model with $r_0=10^{-20}$. 
We observe very clearly the predicted ranges with power-laws $t^2,t^1,t^6$ and, lastly, the diffusive 
$t^1$ range at long times expected for a model with finite $L.$ For comparison, we also plot 
numerical solutions of the mean-field equation (\ref{MF}) using the diffusivity (\ref{KSDT}).  
As before, the mean-field theory predictions are quite close to the full Monte Carlo solution of 
the diffusion model. Lastly, we plot the solution of the mean-field equation for the inertial-range 
large-$L$ diffusivity, with the same choice of constants $L,v_0$ and $\varepsilon.$ 
As expected, the dispersion law from this approximation agrees quite well with that of the 
KS diffusion model for $r<L,$ but predicts a spurious $t^3$ power-law range 
for $r>L.$ The good agreement justifies {\it a posteriori} our simplification of the scaling function
$J(x)$ in section \ref{IRnum}. Our most important general conclusion from this set of experiments 
is that the KS diffusion models and, we believe, the KS models themselves should exhibit 
the above four scaling ranges with successive power-laws $t^2,t^1,t^6$ and then $t^1$ again, 
whenever the scale ratio $k_N/k_1$ is sufficiently large. 

\begin{figure}[!p] 
\begin{center}  
\includegraphics[width=\linewidth,,height=9cm]{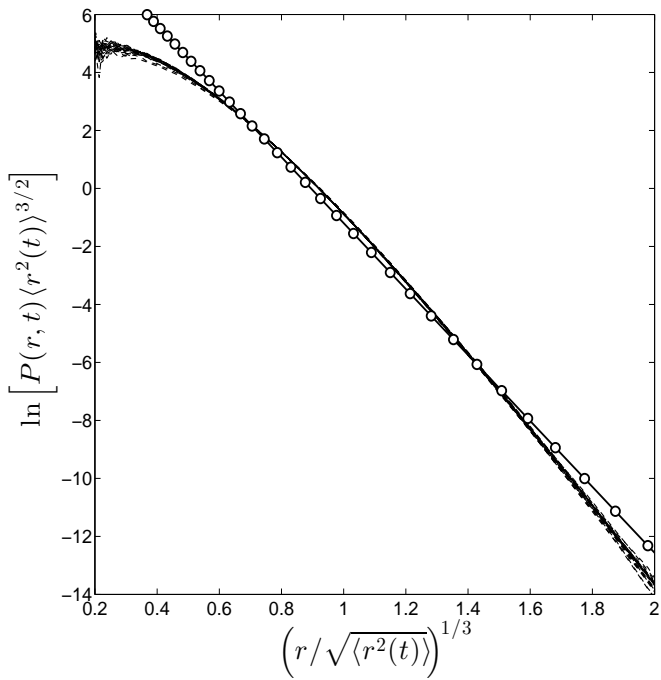}
\caption{Logarithm of the rescaled PDF of pair-separations at $23$ different  times 
in the $t^6$ range, for KS diffusion model with $k_N/k_1=10^{21}$. Monte Carlo 
results for $S=10^6$ and $C_\Delta=1$. The straight line marked with circles
($\circ$) is the analytical result (\ref{SS-PDF}) for $d=3$ and $\gamma=1/3$.}
\label{figure7}
\end{center} 
\end{figure}

In order to discriminate between various alternative theories, it is useful to compare predictions 
not only for mean-square separations but also for the full probability density $P(r,t).$ Although 
we do not expect our adaptive time-stepping algorithm to be sufficient to reproduce accurate 
PDFs, it is still useful to present a few numerical results for the KS diffusion model. In Fig. \ref{figure7} 
we plot the Monte Carlo probability distribution calculated for $36$ different times spread 
within the $t^6$ range.  These are rescaled to test for self-similarity and collapse quite well. 
It should be emphasized that the overall evolution of our IR and KS diffusion models is {\it not}
self-similar,  globally in time. This can be seen most clearly in the existence of time ranges with distinct 
power-law growth laws, whereas a truly self-similar evolution should have just one power-law. 
In a sufficiently long $t^6$ range, however, one should expect a self-similar evolution. For example, the 
inertial-range model (\ref{KL}) in the $t^6$ range reduces to the time-independent diffusivity 
$K_L(r,\infty)=C_L' \varepsilon^{2/3}r^{5/3}/v_0,$ except for $r\gg v_0t.$ Since $r\sim v_0 t$ is 
nearly the maximum particle separation that can be achieved in the time $t,$ only a very tiny 
large-$r$ tail will experience a different eddy-diffusivity than this. In Fig. \ref{figure7}  we 
also compare the Monte Carlo results for the KS diffusion model with the exact parameter-free 
predictions (\ref{alpha}),(\ref{beta}) of the power-law diffusion model (\ref{pow-diff}) for $d=3$ and 
$\zeta=5/3.$ The agreement is reasonably good. Furthermore, the Monte Carlo results for the KS diffusion 
model agree almost perfectly with the Monte Carlo results for the power-diffusion model presented 
in section \ref{NumMeth}. This suggests that if our Monte Carlo could be carried out with a small, 
constant time-step in a long $t^6$-range, then the PDF would approach the exact self-similar form   
of the power-law diffusion model. 

\begin{figure}[!h] 
\begin{center}  
\includegraphics[width=\linewidth,,height=9cm]{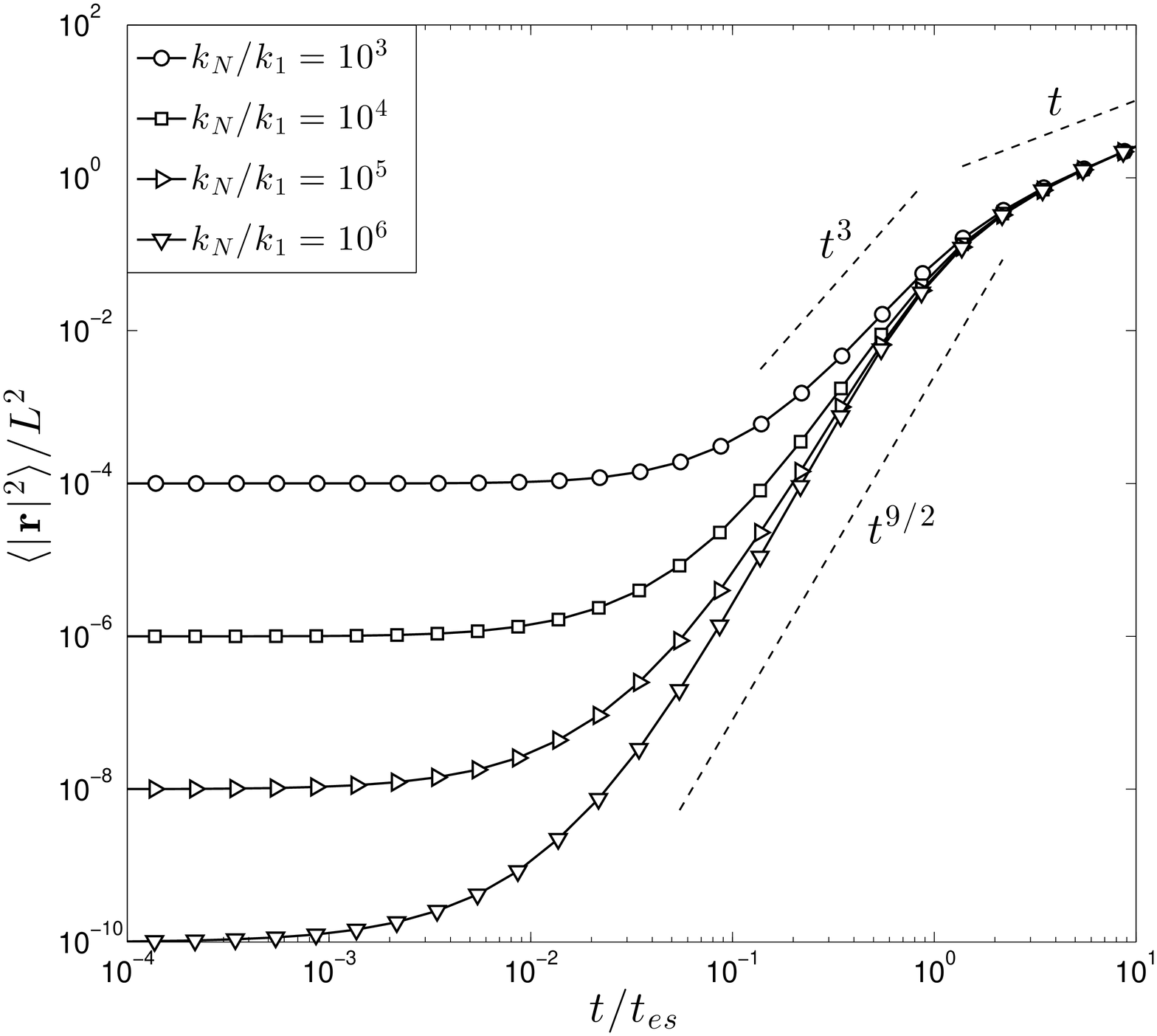}
\caption{Monte Carlo results for $\langle|\mathbf{r}(t)|^2 \rangle$ in KS diffusion model for 
various values of $k_N/k_1$, with $\Delta t=0.1\frac{r}{v_0}$, $S=10^4$.}
\label{figure8}
\end{center} 
\end{figure}

We have conjectured that the growth laws of the KS models themselves, asymptotically for 
$k_N/k_1\gg 1$, are the $t^2,t^1,t^6$ and $t^1$ powers that we have found in the KS diffusion 
models. How can this be reconciled with the $t^{9/2}$ law predicted in \cite{ThomsonDevenish05} and 
verified to greater or lesser extent in subsequent KS simulations \cite{ThomsonDevenish05,
DevenishThomson09,NicolleauNowakowski11}? We argue that the observed $t^{9/2}$ is 
an artefact of the modest $k_N/k_1$ ratios achieved in these simulations, which tends 
to ``blend'' the distinct scaling ranges, in particular the early-time $t^1$ and $t^6$ ranges,
between which lies a broad transition zone. In support of this argument, we have performed 
a sequence of Monte Carlo simulations of the KS diffusion model with scale ratios $k_N/k_1
=10^3,10^4,10^5,10^6$. The last ratio is chosen to correspond roughly to that employed in 
the previous KS simulations \cite{ThomsonDevenish05,DevenishThomson09,NicolleauNowakowski11}.
Because the range of time-scales is not so great, we have been able to carry out the time-integration
not only with the adaptive algorithm employed up until now, but also with a constant time step 
$\Delta t = 0.1 \frac{\eta}{v_0}$
which resolves the effects of even the smallest eddies, equivalent to that used in recent KS simulations
\cite{DevenishThomson09,NicolleauNowakowski11}. The results of the two time-advancement schemes
for the dispersion curves are identical when plotted in log-log. 
As illustrated in Fig.~\ref{figure8}, a $t^{9/2}$ regime seems to appear as we increase the ratio  $k_N/k_1.$  
This figure should be compared with Fig. 2 of \cite{DevenishThomson09} and Fig. 1 of \cite{NicolleauNowakowski11}, 
which it matches very closely. Although we see a similar ``$t^{9/2}$-range'' at the values of $k_N/k_1$ 
used in previous KS simulations, covering 1-2 decades in time, it is clear from our results in Fig.~\ref{figure6} 
that this is only a transitional regime of the KS diffusion model. In fact, for the case $k_N/k_1=10^6$ which
shows the long ``$t^{9/2}$-range'' we find $t_{to}\doteq 10^{-2}$ and thus the broad transition zone 
between the $t^1$ and $t^6$ laws covers the interval from $10^{-4}$ to $10^0.$ This includes all
of the apparent ``$t^{9/2}$-range''. If we go to $k_N/k_1=10^8,$ the power-law steepens into a 
$t^5$-law. At still larger values of $k_N/k_1$ four asymptotic scaling ranges emerge, with distinct 
power-law scalings of $t^2,t^1,t^6$ and $t^1. $ We expect that the same is true of the KS models 
themselves at sufficiently large $k_N/k_1.$

Finally, we note that for $k_N/k_1<10^4,$ the short range of superdiffusive growth of dispersion  
approximates a $t^3$-law. This agrees with the observations of \cite{Osborneetal06,NicolleauNowakowski11}
for KS models. Note, however, that the physics is completely different from turbulent Richardson 
diffusion, which would allow $t^3$ ranges of arbitrary extent. In fact, the narrow range of such a 
power-law in our KS diffusion model arises only because of the ``merging'' of many distinct ranges.
In particular, the exponent of the apparent power-law must decrease with decreasing $k_N/k_1$ to match 
the $t^1$-law starting at $r=L,$ until finally the superdiffusive range disappears entirely when $k_N/k_1\approx 1.$

\section{Conclusions}\lb{Conclude}

We have derived in this paper a diffusion equation for particle-pair dispersion in synthetic
Eulerian turbulence modelled by Gaussian velocity ensembles. The main analytical result 
is the formula (\ref{Gauss-KLform}) for the $2$-particle diffusivity and its special cases (\ref{tauJ}) 
for frozen velocities and (\ref{exact1}) for finite time-correlated velocities.  Although the description 
of pair-dispersion as a diffusion process is not exact (except in certain limiting cases), it arises 
from a well-motivated set of analytical approximations. Our results confirm the physical argument 
of Thomson \& Devenish \cite{ThomsonDevenish05} that pair-dispersion in such models is 
fundamentally altered by sweeping decorrelation effects, not experienced by particle pairs in 
hydrodynamic turbulence. Thus, the $t^3$-law observed in previous simulations with synthetic 
turbulence \cite{ElliottMajda96, FungVassilicos98, MalikVassilicos99, DavilaVassilicos03,
NicolleauYu04} is quite likely an artefact either of the numerical approximations employed or of
the shortness of the inertial ranges. However, we argue as well for a similar origin of the $t^{9/2}$-law  
proposed by Thomson \& Devenish \cite{ThomsonDevenish05} for synthetic turbulence ensembles with 
zero mean velocities. Solutions of our diffusion model for such ensembles at Reynolds numbers 
comparable to those employed in KS simulations that show a $t^{9/2}$-law range reproduce that finding, 
but our model yields instead distinct $t^2,t^1,t^6$ and $t^1$-ranges at higher Reynolds numbers.  
We thus argue that the asymptotic high Reynolds-number behavior of particle dispersion in synthetic 
Eulerian turbulence with zero mean-velocities is the same as that predicted by Thomson \& Devenish 
\cite{ThomsonDevenish05} for ensembles with large mean velocities. 

Synthetic models of turbulence such as Kinematic Simulations have been used to investigate
turbulent transport of passive objects (particles, lines, etc.) in such varied problems as environmental flow, 
aeroacoustics, kinematic magnetic dynamo, and superfluids \cite{Baggaleyetal09,Nicolleauetal11}.
However, such numerical studies must clearly be employed with utmost caution, especially to derive 
conclusions about turbulent transport at very high Reynolds numbers. The difference in sweeping 
effects in synthetic Eulerian turbulence and in real hydrodynamic turbulence imply not only 
quantitatively different scaling laws but also substantially different physics. 

\begin{acknowledgments}
The work of GE was partially supported by the NSF Grant CMMI-0941530 at Johns Hopkins 
University.
\end{acknowledgments}

\appendix
\section{Monte Carlo Time-Step}

\begin{figure} 
\begin{center}  
\includegraphics[width=\linewidth,,height=9cm]{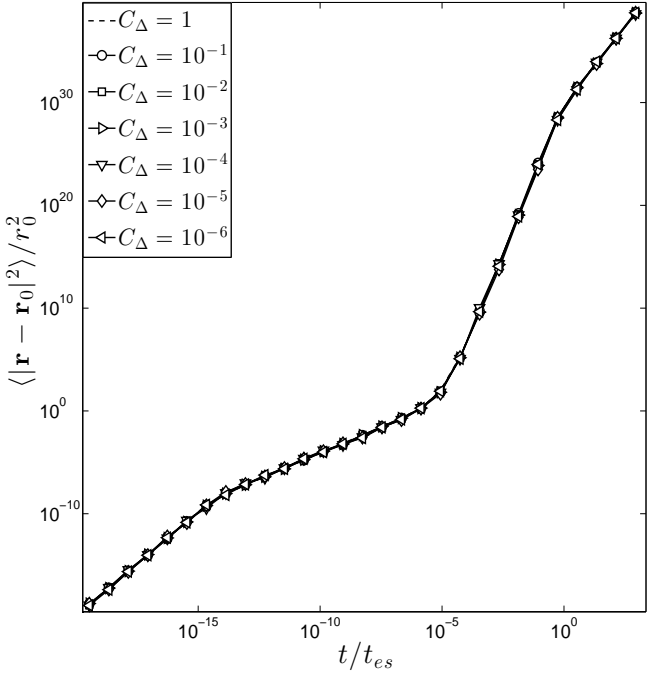}
\caption{Monte Carlo results for $\langle|\mathbf{r}(t)-\mathbf{r}_0|^2 \rangle$ in the inertial-range 
diffusion model calculated with $S=10^2$ samples and varying 
$C_\Delta=1$ to $C_\Delta=10^{-6}.$}
\label{figure9}
\end{center} 
\end{figure}

We tested the dependence of the log-log plots of dispersion on the value of $C_{\Delta}$. 
We plot in Fig. \ref{figure9} the Monte Carlo results for values of $C_{\Delta}$ ranging from 
$1$ to $10^{-6}$. There is no observable change in the behavior. 

\section{Number of Fourier Modes}

\begin{figure}[!h] 
\begin{center}  
\includegraphics[width=\linewidth,,height=9cm]{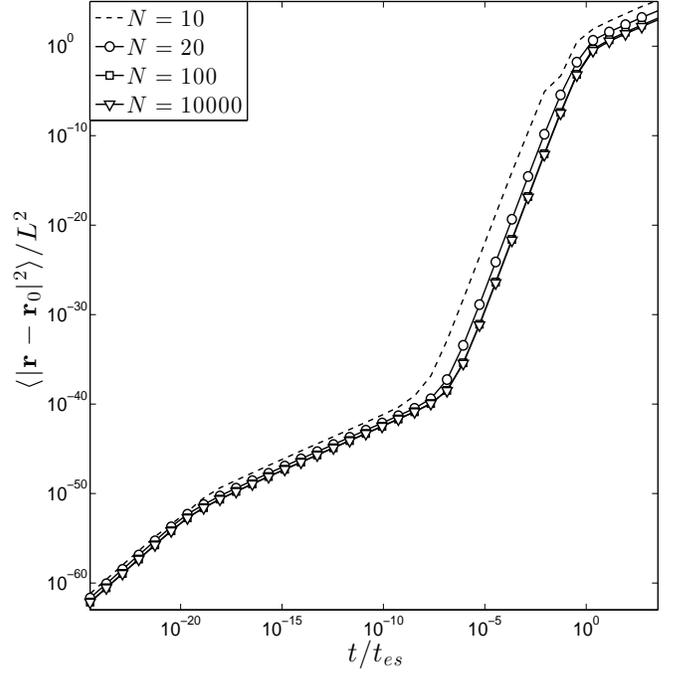}
\caption{Monte Carlo results for $\langle|\mathbf{r}(t)-\mathbf{r}_0|^2 \rangle$ in the KS diffusion model 
calculated with $C_\Delta=1,$ $S=10^4$ samples, varying number of Fourier modes 
from $N=10$ to $N=10^4.$}
\label{figure10}
\end{center} 
\end{figure}

We also tested the dependence of our dispersion results for the KS diffusion models on the number 
of Fourier modes $N$. We show in Fig. \ref{figure10}  log-log plots of the dispersion curves for 
different values of $N,$ obtained from Monte Carlo calculations with $C_\Delta=1$ and $S=10^4$. 
The results are nearly indistinguishable for $N\gtrsim 100$. All of our simulations in the text used $N=500$.

\newpage

\newpage

\bibliography{GaussSweep}

\end{document}